\documentclass[aps,prl,longbibliography,amsmath,amssymb,groupedaddress,superscriptaddress,floatfix,twocolumn,showkeys,nofootinbib]{revtex4-1}
    \usepackage{graphicx}
    \usepackage{dcolumn}
    \usepackage{bm}
    \usepackage{color}
    \usepackage{xcolor}
    \usepackage{multirow}
    \usepackage{siunitx}
    \usepackage{ulem}
    \DeclareSIUnit{\sqrthz}{\ensuremath{\sqrt{\text{\hertz}}}}
    \usepackage{soul}
    \usepackage{makecell}
    \usepackage{svg}
    \usepackage{latexsym}
    \usepackage{amssymb}
    \usepackage{textcomp}

    \usepackage{siunitx}
    \newcommand{\kHz}{\kilo\hertz}
    \newcommand{\MHz}{\mega\hertz}
    \newcommand{\GHz}{\giga\hertz}

    \newcommand{\mW}{\milli\watt}
    \newcommand{\uW}{\micro\watt}
    \newcommand{\nm}{\nano\meter}
    \newcommand{\us}{\micro\second}
    \newcommand{\ms}{\milli\second}
    \newcommand{\mm}{\milli\meter}
    \newcommand{\uK}{\micro\kelvin}
    \newcommand{\mK}{\milli\kelvin}
    \newcommand{\K}{\kelvin}
    \newcommand{\mG}{\milli G}
    \newcommand{\um}{\micro\meter}
    \newcommand{\ket}[1]{\ensuremath{\left| {#1} \right>}}
    \newcommand{\mrad}{\milli\radian}
    \newcommand{\mbar}{\milli\bar}
    \usepackage[T1]{fontenc}

    \makeatletter
    \renewcommand{\c@secnumdepth}{0}
    \makeatother

\begin{document}

\title{A high optical access cryogenic system for Rydberg  \\ atom arrays with a 3000-second trap lifetime}

\author{Zhenpu Zhang}
    \thanks{These two authors contributed equally}
    \affiliation{JILA, National Institute of Standards and Technology and University of Colorado, Boulder, Colorado 80309, USA}
    \affiliation{Department of Physics, University of Colorado, Boulder, Colorado 80309, USA}
\author{Ting-Wei Hsu} 
    \thanks{These two authors contributed equally}
    \affiliation{JILA, National Institute of Standards and Technology and University of Colorado, Boulder, Colorado 80309, USA}
    \affiliation{Department of Physics, University of Colorado, Boulder, Colorado 80309, USA}
\author{Ting You Tan} 
    \affiliation{JILA, National Institute of Standards and Technology and University of Colorado, Boulder, Colorado 80309, USA}
    \affiliation{Department of Physics, University of Colorado, Boulder, Colorado 80309, USA}
\author{Daniel H. Slichter} 
    \affiliation{Time and Frequency Division, National Institute of Standards and Technology, Boulder, Colorado 80305, USA}
\author{Adam M. Kaufman}
    \affiliation{JILA, National Institute of Standards and Technology and University of Colorado, Boulder, Colorado 80309, USA}
    \affiliation{Department of Physics, University of Colorado, Boulder, Colorado 80309, USA}
\author{Matteo Marinelli}
    \affiliation{JILA, National Institute of Standards and Technology and University of Colorado, Boulder, Colorado 80309, USA}
    \affiliation{Department of Physics, University of Colorado, Boulder, Colorado 80309, USA}
\author{Cindy A. Regal}\email{regal@colorado.edu}
    \affiliation{JILA, National Institute of Standards and Technology and University of Colorado, Boulder, Colorado 80309, USA}
    \affiliation{Department of Physics, University of Colorado, Boulder, Colorado 80309, USA}

\date{\today}

\begin{abstract}
We present an optical tweezer array of $^{87}$Rb atoms housed in an cryogenic environment that successfully combines a \SI{4}{\K} cryopumping surface, a \SI{<50}{\K} cold box surrounding the atoms, and a room-temperature high-numerical-aperture objective lens.  We demonstrate a \SI{3000}{\second} atom trap lifetime, which enables us to optimize and measure losses at the $10^{-4}$ level that arise during imaging and cooling, which are important to array rearrangement. We perform both ground-state qubit manipulation with an integrated microwave antenna and two-photon coherent Rydberg control, with the local electric field tuned to zero via integrated electrodes. We anticipate that the reduced blackbody radiation at the atoms from the cryogenic environment, combined with future electrical shielding, should decrease the rate of undesired transitions to nearby strongly-interacting Rydberg states, which cause many-body loss and impede Rydberg gates. This low-vibration, high-optical-access cryogenic platform can be used with a wide range of optically trapped atomic or molecular species for applications in quantum computing, simulation, and metrology.

\end{abstract}

\maketitle

\section{Introduction}
Neutral atoms are an increasingly mature, high-performance platform for quantum science. Recent advances have enabled individual control of trapped particles within increasingly large-scale systems, extending now to a variety of atomic and molecular species~\cite{bakr2009quantum, sherson2010single, endres2016atom, barredo2016atom, norcia2018microscopic, cooper2018alkaline, saskin2019narrow, liu2018building, anderegg2019optical}. A key property of the neutral atoms in these experiments is their isolation from the environment, which allows many-body quantum states to persist on suitably long timescales.
This combination of good coherence and single-atom control has opened new scientific possibilities for neutral-atom-based systems such as on-demand assembly of defect-free quantum systems~\cite{endres2016atom, kumar2018sorting, lee2017defect-free} and time-keeping based on neutral atoms in optical tweezer arrays~\cite{norcia2019seconds, madjarov2019Atomic,young2020half}. These desirable features can be combined with Rydberg-mediated entanglement generation to enable quantum simulation of spin models~\cite{isenhower2010demonstration,scholl2021quantum,ebadi2021quantum} or quantum information processing with dynamically reconfigurable qubit connectivity~\cite{levine2019parallel,madjarov2020high, bluvstein2024logical}.

The rapid progress of the platform is driven by improved understanding of key noise sources, decoherence mechanisms, and challenges to scaling~\cite{saffman2016quantum,de2018analysis,madjarov2020high,evered2023high,manetsch2024tweezer}. 
Among the challenges going forward are the blackbody environment that can drive undesired transitions during Rydberg-mediated operations and the background-gas-limited atom trap lifetime that can limit the maximum number of controllable atoms.
Operation in a cryogenic environment offers a path to overcome both these obstacles.
Trapped ion experiments often operate at cryogenic temperatures to extend the time ions are held without background gas collisions~\cite{dubielzig2021cryo}, to reduce electric field noise~\cite{Labaziewicz2008a, Chiaverini2014, Sedlacek2018a}, and to more quickly cycle the apparatus~\cite{brandl2016cryogenic,pagano2018cryogenic, Todaro2020}. 
For arrays of neutral atoms and molecules in optical traps, there are similarly important implications if one can perform robust optical control in a cryogenic environment.

Firstly, the temperature of the blackbody environment is an important consideration when using Rydberg states, as blackbody radiation (BBR) induces transitions, primarily to nearby states, via the microwave ``tail'' of the blackbody distribution, which is reduced in a cryogenic environment. 
This effect is particularly noticeable in circular Rydberg atoms, whose radiative lifetime in the absence of BBR can reach seconds but is reduced by orders of magnitude in the presence of 300 K BBR~\cite{raimond2001manipulating,nguyen2018towards,cantat2020long,cohen2021quantum}.
Even in low-angular-momentum Rydberg states used in many-body spin model simulations~\cite{semeghini2021probing,zeiher2017realization}, BBR can shorten the overall lifetime~\cite{beterov2009quasiclassical}.
Improving the lifetime to values limited only by spontaneous decay will add to strategies in place to address errors and loss arising from short Rydberg lifetimes, such as increasing Rydberg Rabi frequencies~\cite{madjarov2020high,evered2023high} and erasure conversion ~\cite{wu2022erasure, ma2023high, scholl2023erasure,chow2024circuit}.  
Equally importantly, BBR-induced populations in opposite-parity Rydberg states open strong interaction channels~\cite{boulier2017spontaneous,festa2022blackbody}.  These unwanted interactions cause rapid many-body loss effects that impair the study of spin models~\cite{zeiher2016many,manovitz2025quantum}, and also degrade gate fidelities in large arrays and when operating with high gate depth~\cite{saffman2016quantum,evered2023high}.
Additionally, BBR limits the accuracy of optical clocks through systematic shifts in spectroscopy~\cite{safronova2012blackbody,ushijima2015cryogenic,beloy2014atomic} and drives undesired transitions between rovibrational states in molecular systems~\cite{ni2018dipolar,holland2024demonstration, Liu2024} and from metastable states in alkaline-earth atoms~\cite{takamoto2011frequency, young2020half}.

In addition to reducing BBR, cryogenic systems also lower the background gas pressure through cryopumping.
Due to the limited depth of optical traps, collisions with background gas can impart sufficient kinetic energy to eject a trapped atom.  For quantum information applications, as systems with thousands of atomic qubits become available, background-gas-induced losses limit the ability to perform operations with long durations that require low error rates. 
Advances in continuous operation and mid-circuit atom replenishment~\cite{norcia2024iterative,gyger2024continuous} offer promising solutions, pairing well with cryogenic conditions that provide long-lived atom reservoirs.
Defect-free preparation through rearrangement is limited by both background-gas-induced atom loss~\cite{barredo2016atom,ebadi2021quantum} and losses associated with laser cooling and imaging during the sequence~\cite{tian2023parallel}. Careful implementation of extremely high vacuum (XHV) practices in room-temperature systems have enabled lifetimes of up to \SI{1400}{\second} in optical tweezer arrays~\cite{manetsch2024tweezer}.  For lifetimes greater than \SI{1000}{\second}, imaging losses are expected to limit the size of defect-free arrays achievable through rearrangement algorithms~\cite{tian2023parallel}.  However, as atom arrays approach 10,000 atoms, and imaging loss is improved, longer atom trap lifetimes will be increasingly important.

A high-optical-access system incorporating cryo-pumping is also relevant for other applications of neutral atoms and molecules. For metrology applications, longer trap lifetimes enable longer interrogation times of optical clock transitions, which is increasingly relevant for exploiting advances in clock coherence~\cite{young2020half, bothwell2022resolving,zheng2022differential,clements2020lifetime}.  Further, in ultracold gases, it has also been hypothesized that the minimum achievable temperature of fermionic quantum gases is particularly sensitive to the loss of atoms deep in the Fermi sea, which may result from background gas collisions~\cite{timmermans2001degenerate,ji2024observation}.

\begin{figure}[h!]
    \centering
    \includegraphics[]{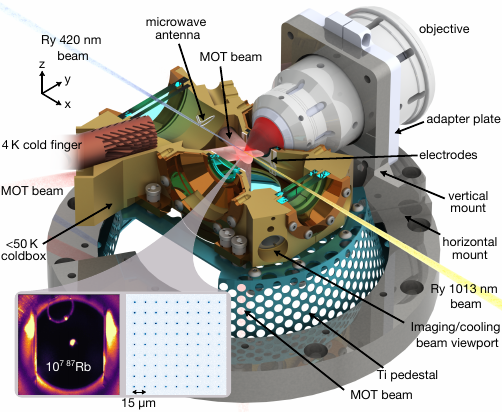}
    \caption{Overview of the core of the cryogenic tweezer apparatus. 
    $^{87}$Rb atoms are trapped in an array of optical tweezers generated by an in-vacuum objective at room temperature. 
    The atom array is surrounded by the cold box at 45-\SI{50}{\K} (yellow volume) and the trapped atoms are controlled with laser beams impinging from various directions. 
    The cold box is mounted on a titanium pedestal for thermal insulation from the room-temperature UHV chamber and incorporates built-in electrodes and a microwave antenna.
    The objective (white) mounts on the adapter plate and vertical and horizontal mount (grey), allowing adjustment for optical alignment.
    The two insets are a camera image of the typical 3D MOT (left) used to load a 2D optical tweezer array (right).
    }
    \label{fig:setup_overview}
\end{figure} 

In this work, we introduce an apparatus that hosts an optical tweezer array in a cryogenic environment with high-numerical-aperture (NA) optical access.  We demonstrate a long atom trap lifetime of \SI{3000}{s} enabled by cryopumping of background hydrogen gas, qubit control using an integrated microwave antenna, and accessibility of Rydberg states through two-photon excitation of the 70$S_{1/2}$ state of $^{87}$Rb.  Previous works have shown trapping and manipulation of single atoms inside a cryostat~\cite{Schymik2021single,pichard2024rearrangement}, but not a comprehensive demonstration of tweezer arrays with long trap lifetimes, large field of view (FoV) with a high-NA lens, and atomic state control, including coherent Rydberg excitation. 

Our concept is based on decreasing the temperature of all surfaces surrounding the atoms to \SI{<50}{\K} with a high cooling power radiation shield that is differentially pumped with respect to an outer ultra-high-vacuum (UHV) region (Fig.~\ref{fig:setup_overview}). Within the radiation shield, a \SI{4}{\K} cryopumping surface that can be periodically regenerated reduces the hydrogen gas partial pressure. Our vacuum conditions are achieved with an in-situ vacuum bake of the radiation shield and \SI{4}{\K} surface to only \SI{90}{\degreeCelsius} and a relatively small \SI{4}{\K} cryopumping surface, which in the future could be naturally extended within our design to subtend a larger solid angle around the atoms. A beam atomic source from a two-dimensional (2D) magneto-optical trap (MOT) enters the radiation shield through a small hole, and multiple stages of differential pumping enable operation with continuous low-NA line of sight to the source.  A key aspect of the atomic control toolbox is optical access for laser beams and high-NA objective. We therefore employ many cold windows on the radiation shield that maintain excellent alignment and flatness during cooldown, in particular with respect to the objective that creates the optical tweezer array.  The in-vacuum objective providing a large FoV is placed outside the radiation shield and remains at room temperature, affording flexibility in objective design.  

Harnessing the long atom trap lifetime in combination with high-performance polarization gradient cooling, we measure and optimize cooling and imaging losses to achieve some of the lowest values reported with alkali atoms~\cite{manetsch2024tweezer,blodgett2023imaging}.  Given these loss rates, we explore the feasibility and current limitations of assembling a large, defect-free array of atoms through Monte Carlo simulation. We use single-qubit microwave rotations to characterize the magnetic field inside the radiation shield, which we find to be stable at the sub-mG level. We demonstrate Rabi oscillations between a ground and Rydberg state, and use electrodes integrated inside the radiation shield to control the static electric field at the atoms. Future iterations of the experiment will incorporate conductive indium tin oxide (ITO) coatings on the windows of the copper radiation shield. The resulting microwave shielding, combined with the cold temperature of the materials surrounding the atoms, should provide substantial suppression of microwave radiation that can cause excitation out of our chosen Rydberg state. Our temperature-tunable radiation shield will aid in characterization of the effectiveness of this suppression~\cite{cantat2020long,norrgard2021quantum} and shielding through studies of interacting Rydberg atoms in this environment.

\section{Apparatus}

\subsection{Cryogenic System Design Concepts}

Despite the long history of cryogenics in a range of neutral atom, trapped ion, and molecule experiments, there have been few cryogenic optical tweezer and quantum gas microscope explorations. This is in part due to the difficulty of combining substantial optical access with cryogenic temperatures and meeting the stringent alignment and flatness requirements for high NA lenses. Recent examples in which optical tweezers were operated in cryogenic environments include Ref.~\cite{Schymik2021single}, which leveraged cryogenic temperature to achieve a \SI{6000}{\second} trap lifetime of single atoms, but with limited optical access, and Ref.~\cite{pichard2024rearrangement}, in which large atom arrays were created, but with line of sight to room-temperature surfaces such that the trap lifetime was substantially reduced.

In this work, we present a cryogenic solution that takes a unique approach to radiation shield design and optical access within a low-vibration system. We construct a custom radiation shield\textemdash which we will refer to as the ``cold box''\textemdash that maintains the required optical alignment and also serves as a differentially-pumped vacuum enclosure. With alignment maintained over a wide range of temperatures, we can trap atoms both with the cryostat on and at room temperature, albeit with reduced trap lifetimes when warm. 
We make use of low-angular-momentum Rydberg states of $^{87}$Rb, which have been used in a range of quantum computing and simulation experiments~\cite{evered2023high, scholl2021quantum}.

The cold box is designed to achieve a base temperature in the range of 40-\SI{50}{\K}, which can be realized without nested radiation shielding and provides substantial expected reduction of Rydberg decay for our target states.  
For example, in the $70S_{1/2}$ state of $^{87}$Rb, where the most relevant transition frequency to nearby Rydberg states is approximately \SI{10}{\GHz}, the BBR-induced decay to nearby Rydberg states (which cause rapid many-body loss and gate or simulation errors when populated) decreases by a factor of 8 in an ideal \SI{40}{\K} blackbody environment compared to room temperature. 
The overall calculated Rydberg lifetime increases to \SI{310}{\us} as a result, compared to the total room-temperature radiative lifetime of \SI{150}{\us} and the zero-temperature radiative lifetime of \SI{370}{\us}.
In addition to the cold box temperature, the emissivity of the constituent materials and their microwave electrical properties are key to achieving the desired environment for Rydberg atoms. 
The copper cold box ideally provides shielding against electric field noise and serves as a barrier to room-temperature BBR at microwave frequencies due to its high electrical conductivity.
The windows are designed to be compatible with electrically conductive ITO coatings for full shielding from room-temperature microwave BBR, enabling future studies of Rydberg decay as a function of cold box temperature.  
The atomic array is placed \SI{8}{\mm} from the inner window surface, a typical distance for Rydberg array experiments, and in future detailed studies we anticipate the opportunity to study the impact of temperature on window-surface electric field drift and noise by measuring Rydberg atom coherence~\cite{mamat2024mitigating,thiele2014manipulating}.
We note that the microwave mode structure defined by the electrical boundary conditions of the cold box and temperature range of our system were not optimized for use with long-lived circular Rydberg states~\cite{cantat2020long}.  

Cryopumping of hydrogen is accomplished by a \SI{4}{\K} cold finger housed within the cold box.  To simplify the process of refreshing the cryopumping surface we do not use charcoal sorbent. The cold box itself provides a low pressure environment, due to the negligible line of sight from the atoms to the outer vacuum space (whose pressure is dominated by hydrogen-outgassing stainless steel)~\cite{vittorini2013modular}, and because it is cold enough to cryopump most other background gases. Fully enclosing the atoms with a \SI{4}{\K} box would block hydrogen line of sight to the atoms, resulting in more complete cryopumping, but would complicate optical access, and hence we use only a relatively small \SI{4}{\K} surface in the work demonstrated here.

\subsection{Cryostat and Vacuum Chamber}\label{cryostat_and_vacuum_chamber}

\begin{figure*}[th]
    \includegraphics[]{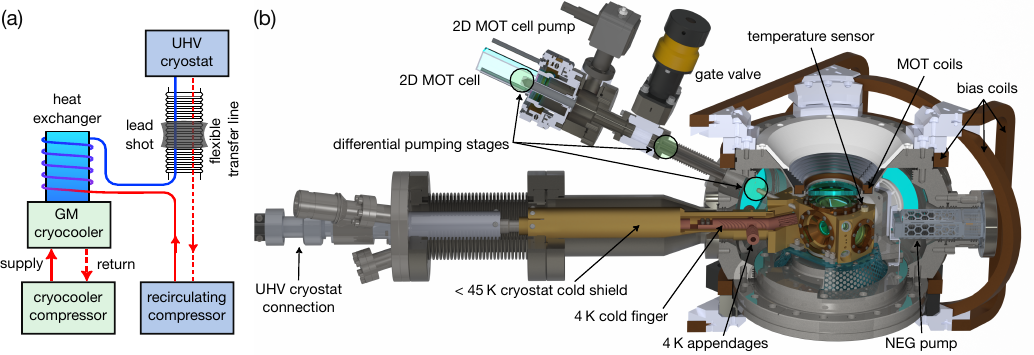}
    \caption{Mechanical details of the cryogenic tweezer apparatus. 
    (a) Schematic of the closed-cycle helium cooling system.
    The first helium loop is connected to a GM cryocooler (green blocks) and provides the primary cooling power.
    The second helium loop, driven by a recirculating compressor (blue blocks), is thermally coupled to the cold stages of the GM cryocooler.
    The cold helium is directed to the gas-flow cryostat via a flexible helium transfer line.
    The flexible helium transfer line is vibrationally dampened by packs of lead shot.
    (b) Mechanical assembly of the UHV cryostat and main vacuum chamber.
    The cryostat includes a \SI{4}{\K} cold finger and a less than \SI{45}{\K} cold shield.
    Two \SI{4}{\K} cryopumping appendages extend outside the cold shield into the main chamber to reduce the vacuum pressure outside the cold box. 
    The image also highlights three differential pumping stages (circles) between the 2D MOT and the interior of the cold box.
    The 3D MOT coils (top coil seen in cutaway view, bottom coil not visible) are recessed and housed within reentrant viewports.
    The cartridge on the right is one of the two NEGs; the ion pump and the other NEG are not visible in this schematic.
    A diode temperature sensor is affixed to the top right corner of the cold box (beige cylinder).
    The temperature sensors for the cold finger and cold shield are not visible in this figure.
    A comprehensive set of images representing the setup is available in Ref.~\cite{hsu2024high}.
    }
    \label{fig:mechanical_detail}
\end{figure*} 

We use a closed cycle system for cryogenic cooling, consisting of three main parts: two separate recirculating helium loops and a UHV-compatible gas-flow cryostat inserted into the vacuum chamber. The first loop operates a typical two-stage Gifford-McMahon (GM) cryocooler [green blocks in Fig.~\ref{fig:mechanical_detail}(a)]. The second helium loop [blue blocks in Fig.~\ref{fig:mechanical_detail}(a)] is driven by a recirculating compressor, which flows gaseous helium through multiple stages of heat exchangers attached to the cold stages of the GM cryocooler and cools it to roughly \SI{4}{K} (Appendix~\ref{Appendix:Cryocooler}). The cold helium then continues to the gas-flow cryostat via a flexible vacuum-jacketed helium transfer line; return flow travels back to the recirculating compressor [via additional cold heat exchangers, not shown in Fig.~\ref{fig:mechanical_detail}(a) for clarity]. The flexible line is designed to decouple vibrations generated by the GM cryocooler from the optical table and vacuum setup, and relax space constraints near the vacuum chamber. 

The main vacuum enclosure is a standard stainless steel chamber with conflat (CF) flanges. The gas-flow cryostat enters the chamber through a bellows to facilitate small position adjustments [left portion of Fig.~\ref{fig:mechanical_detail}(b)]. 
The magnetic field coils for the three-dimensional (3D) MOT are placed on two custom DN160 re-entrant viewports, outside of the vacuum chamber, yet within \SI{40}{\mm} of the MOT and tweezer array locations. 
The atom source is a commercially available 2D MOT in a glass cell, connected to the main chamber with a short vacuum bellows, pumped with a \SI[per-mode = symbol]{3}{\liter\per\second} ion pump referred to as the 2D MOT cell pump [top left of Fig.~\ref{fig:mechanical_detail}(b)]. 
Two differential pumping stages isolate the 2D MOT cell from the rest of the vacuum system, with an approximate conductance of \SI[per-mode = symbol]{0.05}{\liter\per\second} and \SI[per-mode = symbol]{0.02}{\liter\per\second} between the 2D MOT and the main chamber [top left of Fig.~\ref{fig:mechanical_detail}(b)].
Throughout this manuscript, the conductance values are referenced to nitrogen at room temperature. 

The main vacuum chamber is pumped with a \SI[per-mode = symbol]{20}{\liter\per\second} (for nitrogen) ion pump [Gamma Vacuum 25S TiTan, not visible in Fig.~\ref{fig:mechanical_detail}(b)] and two SAES non-evaporable getter (NEG) cartridges [one cartridge (NexTorr Z200) is on the right in Fig.~\ref{fig:mechanical_detail}(b); the other NEG (Capacitorr Z200) is not visible]. 
The integrated ion pump within the NexTorr Z200 is turned off and not shown, with no role in the experiment. Throughout this manuscript, the ``ion pump'' refers specifically to the Gamma Vacuum pump in the main chamber.
The NEGs are directly mounted to the vacuum chamber via DN40 CF flanges, while the ion pump is connected to the chamber through a \SI{76}{mm} bellows attached to a DN40 CF flange.
During NEG activation, the cold box and cryostat are heated to \SI{90}{\degreeCelsius} (a temperature well below the melting point of the indium seals used on the cold box windows) for several days using built-in heaters to reduce the chance that molecules outgassing from the NEG will stick to the cold box or cryostat surfaces.
After NEG activation, the ion pump, in the outer vacuum region, measures a pressure of less than \SI{5e-11}{\mbar} at room temperature.
In addition to the ion pumps and NEGs, the outer vacuum is further improved through cryopumping from two \SI{4}{\K} appendages [Fig.~\ref{fig:mechanical_detail}(b)] that protrude out of the cold shield into the main chamber. 
The appendages are chosen to be small so that their blackbody thermal load does not impact the base temperature of the cold finger. 
When the cryostat is cold, the ion pump current corresponds to a nitrogen-calibrated pressure below \SI{1e-11}{\mbar}, which is the noise floor.
We expect the pressure at the atoms to be relatively decoupled from the outer vacuum pressure compared to other factors, as discussed below.

\subsection{Cold Box and Optics}\label{cold_box_and_optics}
The cold box is constructed from a combination of aluminum bronze and oxygen-free-high-conductivity copper (OFHC), with a slit cut to reduce eddy currents caused by magnetic field changes during the experimental sequence. 
It is rigidly mounted to the room-temperature chamber using a pedestal made of perforated titanium sheet bent and welded into a truncated cone (turquoise part in Fig.~\ref{fig:setup_overview}). 
Small polyether ether ketone (PEEK) spacers between the pedestal and cold box provide additional thermal insulation. 
The cold box is attached to the gas-flow cryostat's cold shield with a short copper braid [shorter than \SI{40}{\mm}, not visible in Fig.~\ref{fig:setup_overview} or Fig.~\ref{fig:mechanical_detail}(b)] that provides additional vibration decoupling.

Inside the cold box are a pair of alumina printed circuit boards, with four silver traces each, to control the electric field at the atoms and a resonant microwave antenna for coupling \SI{6.8}{\GHz} microwave fields (Fig.~\ref{fig:setup_overview}). 
The microwave magnetic fields are used to drive the ground-state hyperfine transitions in $^{87}$Rb while the electric fields are used for Rydberg state readout by selectively removing Rydberg atoms.  
The antenna is a \SI{6}{mm}-diameter loop formed at the end of a rigid coaxial cable, with a small cut in the loop opposite the coaxial cable (see Sec.~\ref{sec:Ryberg_excitation} and~\ref{Microwave_qubit}).  Both the dc electrodes and the microwave antenna are connected to cryogenic filters with appropriate passbands that are thermalized to the cold box temperature to block room-temperature BBR.

The cold box is enclosed with BK-7 windows coated with ion beam sputtered (IBS) anti-reflection coatings and attached with indium gaskets (200-\SI{300}{\um} thick) on both sides to facilitate thermal contact and relieve stresses induced by differential thermal expansion (Sec.~\ref{cryogenic_hardware_perfomance}).
The two windows of the cold box that can interface with the high-NA objectives (recessed windows in Fig.~\ref{fig:setup_overview}) are aligned with a Fizeau interferometer to be parallel within \SI{0.1}{\mrad} with a peak-to-valley flatness within $\lambda/5$, for a $\lambda$ of \SI{633}{\nm}.
Besides the two high-optical-quality windows used for high-NA tweezer beams, the cold box is equipped with four additional viewports of larger size, designed for optical lattice beams and excitation beams, such as those for optical pumping, Rydberg excitation, and Raman sideband cooling.
Each of these viewports features double-pane construction (Fig.~\ref{fig:setup_overview}) to reduce the blackbody radiation load on the inner window to better thermalize it to the temperature of the cold box.

The cold box is designed to interface with up to two room-temperature high-NA objectives in a confocal arrangement, although only one is used in the current setup.
The objective is a custom design with an NA of 0.55 (Special Optics\footnote{Certain equipment, instruments, software, or materials are identified in this paper in order to specify the experimental procedure adequately.  Such identification is not intended to imply recommendation or endorsement of any product or service by NIST, nor is it intended to imply that the materials or equipment identified are necessarily the best available for the purpose.}), a working distance of \SI{14}{\mm} (comprising \SI{3}{\mm} of vacuum outside the cold box window, a \SI{3}{\mm} thick BK-7 window, and \SI{8}{\mm} of vacuum inside the cold box), a FoV of \SI{>400}{\um} diameter with a Strehl ratio greater than 0.8, and a chromatic focal shift of less than \SI{0.2}{\um} from 750-\SI{1100}{\nm}. 
We have also successfully operated with a large single aspheric lens instead of the objective, and the image in Fig.~\ref{fig:setup_overview} was made using this lens; all other data in this manuscript use the custom objective.
The objective mount consists of three parts: the vertical mount, the horizontal mount, and the adapter plate (Fig.~\ref{fig:setup_overview}), all of which assure precise alignment and relative positional stability of the objective to the cold box while establishing a rigid connection to the main chamber. The monolithic horizontal mount is designed to reduce differential vibrations in anticipation of mounting two objectives in a confocal configuration.
Angular alignment is achieved during assembly via shimming of the vertical mount, after which the system is secured with screws to maintain positional integrity.

The cold box is equipped with a differential pumping tube with a conductance of around \SI[per-mode = symbol]{0.03}{\liter\per\second} [Fig.~\ref{fig:mechanical_detail}(b)] that provides an opening for the atom beam.
This opening also provides a line of sight for gas molecules from the outer vacuum to collide with trapped atoms inside the cold box. 
Based on the solid angle subtended by the opening and the outer vacuum pressure, we estimate this effect to contribute negligibly to our measured atom trap lifetime.
In addition to the atom source opening, there are two other pathways coupling the outer vacuum space to the cold box interior. 
One is through the region where the cold box connects to the cryostat cold shield, which contains openings for the two \SI{4}{\K} appendages, with an estimated conductance below \SI[per-mode = symbol]{4}{\liter\per\second}. 
The second is the slit on the cold box, with an estimated conductance of less than \SI[per-mode = symbol]{0.1}{\liter\per\second}. 
Importantly, none of these conductance paths provide a direct line of sight to the atoms from the outer vacuum. 
This ensures that gas entering the cold box collides with at least one surface below \SI{50}{\K}, with gas near the cryostat cold shield openings likely cryopumped by a \SI{4}{\K} surface before reaching the atoms.

\subsection{Cryogenic Hardware Performance}\label{cryogenic_hardware_perfomance}
\begin{figure*}[th]
    \includegraphics[]{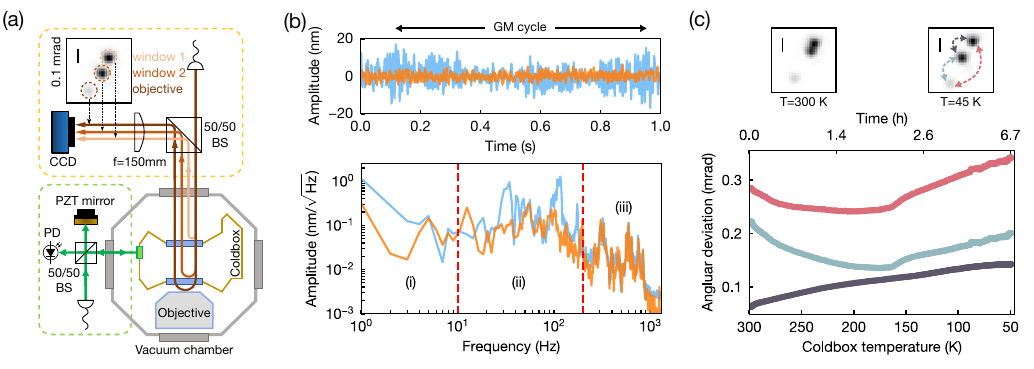}
    \caption{Cold box vibration measurements and characterization of optical alignment during cooldown. 
    (a) Michelson interferometer used to measure the relative vibrations between the cold box and the optical table (green dashed box). 
    Setup monitoring the beam reflections from the cold box windows and objective to monitor angular deviations during a cooldown (yellow dashed box). 
    (b) Vibration spectrum recorded by the Michelson interferometer. 
    The traces are measured with (orange) and without (blue) packs of lead shot placed on the flexible helium line [see Fig.~\ref{fig:mechanical_detail}(a)]. 
    The lead shot dampens vibrations up to \SI{\sim 200}{\hertz}, originating from the GM cycle frequencies (region i) and mechanical resonances of optical table and mounts that are excited by the GM (region ii). 
    (c) Relative alignment of the cold box windows and the objective during cooldown, including the objective to the nearer window (blue-grey), the objective to the farther window (light red), and between the two windows (dark violet). 
    The measured angular misalignment remains within the tolerance of the objective design (\SI{1.5}{\mrad}) for achieving diffraction-limited spot sizes. 
    The scale bar in the small panels indicates an angular deviation of \SI{0.1}{\mrad}.
    }
    \label{fig:vibration_charac}
\end{figure*} 

We monitor the temperature of the cold finger and cold box using in-situ silicon diode thermometers.
All temperature measurements of the cold box are made on the metal box [small beige cylinder on the top right corner of the cold box in Fig.~\ref{fig:mechanical_detail}(b)], and temperatures at the center of the dielectric windows may be higher.  
The cold finger (cold box) cools to its base temperature of \SI{4}{\K} (\SI{45}{\K}) in 2 hours (8 hours). 
This cold box temperature is achieved when the system is freshly cooled from room temperature.
Over several weeks, it gradually rises to around \SI{50}{\K} due to reduced cooling power caused by partial clogging of the recirculating helium circuit for the gas-flow cryostat, while the cold finger remains at \SI{4}{\K} due to its minimal thermal load. 
Similarly, the cold finger and associated appendages can be heated independent of the cold box, and this capability is used to refresh the cold surfaces daily (Appendix~\ref{Appendix:regeneration}).

The cooling power of the cryostat is measured by heating the \SI{4}{\K} stage and the cryostat cold shield separately with in-situ heaters in a standalone test chamber and recording temperature as a function of applied heat load.
The thermal load on the cold box is derived by comparing the cryostat cold shield temperature when the cold box is connected to the temperature-heat load curve.
The cooling power of the \SI{4}{\K} stage exceeds \SI{650}{\mW} at \SI{4.2}{\K}, and that of the cryostat cold shield exceeds \SI{4}{\watt} at \SI{40}{\K}. 
The intrinsic heat load on the cold box is determined to be around \SI{3.3}{\watt} and is consistent with a thermal finite element simulation that includes both the blackbody load on the cold box and physical thermal links.

To measure the vibration level, a small mirror is attached to the cold box to form a Michelson interferometer between the cold box and the optical table [green dashed box in Fig.~\ref{fig:vibration_charac}(a)]. 
The vibration amplitude is calibrated using the Michelson fringe spacing and recorded as a time trace [top panel of Fig.~\ref{fig:vibration_charac}(b)].  
The vibration power spectral density reveals three distinct frequency regions [bottom panel of Fig.~\ref{fig:vibration_charac}(b)]: (i) vibrations from the GM cycle (around \SI{2}{\hertz}), (ii) mechanical resonances of components such as the optical breadboard, mirror mounts, and the cold box, and (iii) acoustic vibrations from the laboratory environment. 
Initially, without damping the helium flex line, the root mean square (RMS) vibration amplitude is as low as \SI{7}{\nm}, although excursions from the GM impulse are still visible (blue trace). 
Damping the flexible line with bags of lead shot eliminates these excursions, reducing the measured vibration amplitude to \SI{3}{\nm} RMS, similar to the background level when the cryocooler is off.

The relative optical alignment upon cooldown is assessed by recording the spot position of reflections from three different planar optical surfaces (two from the cold windows and one from the front side of the objective) as a function of cold box temperature, using a collimated incoherent light source launched through the side of the cold box opposite to the objective [Fig.~\ref{fig:vibration_charac}(a)].
Upon cooldown, we observe a change of less than \SI{0.1}{\mrad} in the relative alignment of the two tweezer windows and less than \SI{0.05}{\mrad} change between the objective and the cold box tweezer windows [Fig.~\ref{fig:vibration_charac}(c)]. 
The angular deviation of \SI{0.35}{\mrad} between the objective and the cold box when cooled to less than \SI{50}{\K} is well within the predicted \SI{1.5}{\mrad} tolerance for diffraction-limited performance. 
This alignment change during cooldown is repeatable and consistently settles to the same position for a given cold box temperature.  
Sec.~\ref{Optical_tweezer_lifetime} gives further details of optical performance as measured by the atomic array.

\section{Optical tweezer and atom trap lifetime}  \label{Optical_tweezer_lifetime}

\subsection{Science MOT and Optical Tweezers}
The experimental sequence starts with the loading of atoms into 2D and 3D MOTs.
Atoms pre-cooled in the 2D MOT are directed toward the center of the cold box through the three differential pumping stages and the gate valve [Fig.~\ref{fig:mechanical_detail}(b)].  
We use a far-detuned dipole push beam (\SI{8}{\mW}, with a $1/e^2$ Gaussian beam waist of \SI{0.3}{\mm}, detuned \SI{520}{\MHz} below the $F=2\leftrightarrow F^{\prime}=3$ $D_2$ transition) to achieve a guiding effect that mitigates the challenges of aligning the differential pumping stages to follow the ballistic trajectory of the atoms under the influence of gravity~\cite{ram2011note,bruneau2014guided}. 
We observe that a far-detuned beam results in a higher atom flux at the 3D science MOT location than typical near-resonant push configurations. 

The science MOT within the cold box is generated by three pairs of retro-reflected beams (Fig.~\ref{fig:setup_overview}).
These beams have a waist of \SI{2.5}{\mm}, a power of \SI{4}{\mW} per beam, and are red-detuned by $2\,\Gamma$ from the $F=2\leftrightarrow F^{\prime}=3$ $D_2$ transition, where $\Gamma=2\pi \times \SI{6}{\MHz}$ is the natural linewidth.
The repumping beams, propagating in the same fibers as the MOT beams, are resonant with the $F=1\leftrightarrow F^{\prime}=2$ $D_2$ transition and have a power of \SI{0.2}{mW} per beam.
The magnetic field gradient for MOT loading is \SI{10}{G\per\cm} and the steady-state MOT contains approximately $10^7$ $^{87}$Rb atoms.

The MOT can also serve as a diagnostic tool for evaluating vacuum pressure~\cite{willems1995creating} independent of tweezer lifetime measurements.
By reducing the total power of each MOT beam to around \SI{1}{\mW}, the atom number is decreased to only several hundred, thereby minimizing the effects of light-assisted collisions~\cite{weiner1999experiments} that cause additional losses.
In this configuration, we measure a $1/e$ MOT lifetime of 1 hour with the 2D MOT gate valve closed [Fig.~\ref{fig:mechanical_detail}(b)].  
Given the larger trap depth of a MOT, MOT lifetimes are typically larger than the corresponding optical tweezer trap lifetime.

In the absence of the push beam and 2D MOT, the loading rate of the low-density 3D MOT characterizes the $^{87}$Rb background flux along the atomic beam path. 
The background flux is determined to be on the order of $10^2$ atoms per second, much lower than our typical 3D MOT loading rate of \SI{4e6}{} atoms per second when the atom beam is turned back on.
In Sec.~\ref{sec:vacuum_loss}, we observe that the background flux has no measurable effect on the trap lifetime of atoms in optical tweezers at the few-thousand-second scale.

After loading the MOT for \SI{500}{\ms}, an array of optical tweezers with an average waist of \SI{0.72}{\um} is turned on at a trap depth of $U_0/k_B=$ \SI{0.84}{\mK}.
This is followed by \SI{25}{\ms} of red-detuned polarization gradient cooling (RPGC)~\cite{dalibard1989laser} (\SI{80}{MHz} red-detuned from the trap-light-shifted $F=2\leftrightarrow F^{\prime}=3$ $D_2$ transition) for cooling and probabilistic loading of single atoms into the tweezer potentials. 
As an alternative to RPGC, during the loading stage we use $\mathrm{\Lambda}$-enhanced gray molasses (LGM) in the same beam geometry~\cite{brown2019gray} and achieve single atom loading probabilities exceeding \SI{80}{\percent} across all tweezers.
The optical tweezer array is generated with \SI{830}{\nm} light from a diode-seeded tapered amplifier, deflected into multiple beams by a two-axis acousto-optic deflector (AOD) (AA optoelectronics DTSXY-400-850.930-002). 
The deflected light goes through a series of telescopes and a dichroic beam splitter before being focused through the high-NA in-vacuum objective. 
The presence of single atoms in each optical tweezer is verified by standard fluorescence imaging in which photons scattered by the atoms are collected by the in-vacuum objective and imaged onto a scientific CMOS camera.

The typical tweezer array for the work in this paper consists of a $5\times5$ grid (Fig.~\ref{fig:beam_waist}) with non-uniform spacing, except for the Rydberg excitation in Sec.~\ref{sec:Ryberg_excitation}.
The tweezer spacing is chosen to reduce the impact of intermodulation ~\cite{endres2016atom}.
We perform trap balancing procedures similar to those outlined in Ref.~\cite{jenkins2022ytterbium} and achieve approximately \numrange{5}{10}\SI{}{\percent} peak-to-peak variation in trap depths.
The radial trap frequency is measured using parametric heating, yielding an average value of $\omega_r= 2\pi \times$\SI{123}{\kHz}. 
Based on this frequency and the trap depth, we deduce the NA for the trapping light to be $0.49$.
This result highlights the robustness of the relative alignment between the objective and the cold window at cryogenic temperatures.

\subsection{Tweezer Trap Lifetime and Loss Channels}
\begin{figure*}[th]
    \includegraphics[]{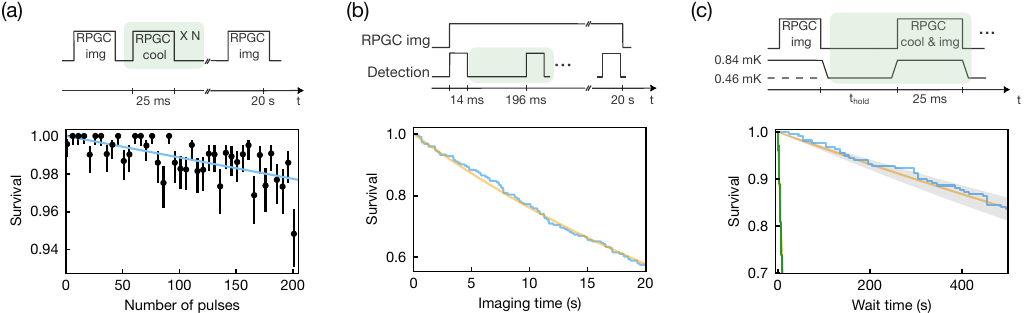}
    \caption{Characterization of atom losses induced by imaging light, cooling pulses and background gas collisions. 
    (a) The cooling loss measurement sequence (top panel) consists of a variable number $N$ cooling pulses (green shaded area), each \SI{25}{\ms} long, evenly spaced over a fixed \SI{20}{\second} interval. 
    The bottom panel shows atom survival as a function of $N$. 
    MLE yields a single-pulse cooling loss probability per atom of \num{1.1(1)e-4} or equivalently a loss rate of \SI{4.4(4)e-3}{\per\second}, with the loss probability depicted by a decaying exponential (blue line) as a visual guide. 
    Error bars on atoms' survival throughout this work represent equal-tailed Jeffrey's prior confidence intervals~\cite{brown2001interval}.
    (b) The sequence to measure imaging loss (top panel) uses RPGC imaging light continuously applied to the atoms while camera exposures of \SI{14}{\ms} duration are taken every \SI{196}{\ms} (green shaded area) for \SI{20}{\second}. 
    The bottom panel shows atom survival as a function of imaging light duration (blue line). 
    MLE yields a loss rate of \SI{2.7(3)e-2}{\per\second} or alternatively \num{3.8(4)e-4} per image duration (\SI{14}{ms}), shown by the orange line as a visual reference. 
    (c) The sequence to measure the loss due to background gas collisions under both cryogenic and room-temperature conditions is shown in the top panel. 
    For cryogenic operation, cooling and imaging pulses are applied every \SI{10}{\second} (green shaded area); for room-temperature measurements, this interval is reduced to \SI{1}{\second} to accommodate higher loss rates.
    In both cases, each cooling and imaging pulse lasts \SI{25}{\ms}, with the trap depth reduced to \SI{0.46}{\mK} during hold time $t_{\text{hold}}$ to reduce parametric heating. 
    The bottom panel shows atom survival as a function of time for cryogenic (blue) and room-temperature (green) conditions. 
    MLE yields an atom trap lifetime of $2800^{+500}_{-400}$ \SI{}{\second} (cryogenic) and $30^{+6}_{-5}$ \SI{}{\second} (room temperature). 
    Solid orange curves and shaded regions show exponential decays with time constants and uncertainties, respectively, as a guide to the eye. 
    After accounting for the \num{2e-4} loss due to cooling and imaging applied every \SI{10}{\second} in the cryogenic condition, the corrected vacuum-limited atom trap lifetime is determined to be $3000^{+600}_{-500}$ \SI{}{\second}.
    }
    \label{fig:loss_charac}
\end{figure*} 
\subsubsection{Cooling and Imaging Loss}\label{sec:cooling_and_imaging_loss}
To determine the vacuum-limited lifetime of atoms trapped in optical tweezers, it is crucial to calibrate and understand all loss channels.  
These include not only background gas collisions but also any processes by which the atoms are removed from the trap, including those occurring during imaging or laser cooling.

One source of loss is the heating of the atom motion either through recoil due to off-resonant scattering or parametric heating due to the intensity noise from the trapping light.
Furthermore, the finite depth of the trap combined with the thermal distribution of the atoms' kinetic energy can lead to atom loss well before the average kinetic energy becomes comparable to the trap depth~\cite{tuchendler2008energy}.
The expected recoil heating rate at a trap depth of $U_0/k_B=$ \SI{0.84}{\mK} is calculated to be \SI[per-mode = symbol]{9}{\uK\per\second}, consistent with the experimentally determined value within the measurement uncertainty.
To reduce parametric heating, the relative intensity noise of the tweezer light is suppressed to below \SI{-120}{dBc\per\hertz} near twice the radial and axial trap frequencies, limiting its contribution to less than \SI[per-mode = symbol]{1}{\uK\per\second}.

To mitigate heating mechanisms during the long sequence used to measure the atom lifetime, we apply RPGC intermittently.
However, it is important to note that cooling processes can also introduce losses, as observed in alkali-metal~\cite{blodgett2023imaging, ang2022gray, tian2023parallel,manetsch2024tweezer, hutzler2017eliminating} and alkaline-earth~\cite{jenkins2022ytterbium, covey20192000} atoms with various cooling and imaging schemes.
Thus, an optimal balance between cooling and heating is required to minimize their combined effect relative to background-gas-induced losses.

In our experimental setup, both imaging and cooling processes employ in-trap RPGC at \SI{0.84}{\mK} trap depth, differing primarily in the number of scattered photons during each operation (Appendix~\ref{Appendix:image_loss}).
Using the release-and-recapture technique~\cite{tuchendler2008energy} with the assumption of a harmonic trapping potential, the temperature after cooling and imaging is measured to be around \SI{10}{\uK}.

The cooling light uses a pair of retro-reflected beams with a waist of \SI{1.1}{\mm} in the RPGC configuration, delivered through a dedicated viewport (Fig.~\ref{fig:setup_overview}) and oriented to project on all three axes of the tweezer trap.
The cooling light is detuned \SI{85}{\MHz} below the trap-light-shifted $F=2\leftrightarrow F^{\prime}=3$ $D_2$ resonance with \SI{1.1}{\mW} power per beam. 
The repumper operates at approximately \SI{1}{\uW} per beam to minimize cooling loss~\cite{schymik2022scaling}, while maintaining cooling efficiency.

Cooling loss is measured using a sequence where the total hold time was fixed at \SI{20}{\second} while varying the number of cooling pulses, each lasting \SI{25}{\ms} [Fig.~\ref{fig:loss_charac}(a)].
Using maximum likelihood estimation (MLE), we determine the cooling loss per \SI{25}{\ms} pulse to be \num{1.1(1)e-4}.
This loss per cooling pulse is approximately half of the smallest \SI{14}{\ms} pulse imaging loss (Appendix~\ref{Appendix:image_loss}), despite its longer duration, with further reduction expected by optimizing cooling time without compromising effectiveness.
Throughout this manuscript, unless otherwise specified, uncertainties in the fitted results represent the bias-corrected and accelerated intervals obtained from bootstrapping~\cite{efron1994introduction}, and all uncertainties represent \SI{68}{\percent} confidence intervals.

The imaging light uses the same beam geometry used for cooling but with a detuning of \SI{90}{\MHz} below the trap-light-shifted resonance, with each beam carrying a power of \SI{2.4}{\mW} and with the repumper saturating the transition.
During imaging, the camera is exposed for \SI{14}{\ms} and detects an average of approximately $220$ photons per trap site (Appendix~\ref{Appendix:image_process}), which yields a discrimination infidelity below \num{1e-6}.

To quantify imaging losses, the imaging light is turned on for \SI{20.144}{\second}, with the camera gated for exposure every \SI{196}{\ms} [Fig.~\ref{fig:loss_charac}(b)].
Given the longitudinal nature of the data acquisition, the assumption of independence between points in the survival versus time trace is no longer valid. 
Therefore, we employ MLE with censoring~\cite{klein2003survival} to extract the exponential decay constant: the fitted imaging loss is \SI{2.7(3)e-2}{\per\second} or equivalently, \num{3.8(4)e-4} loss per \SI{14}{\ms} image.
This loss is comparable to the best-reported loss of \num{1.05e-4} per image measured with Cs~\cite{manetsch2024tweezer}. 
However, our imaging time is about 6 times shorter than the \SI{80}{\ms} reported in that study. 
A shorter imaging time is advantageous for fast readout in multi-round rearrangement sequences and for other applications such as quantum error correction.
With our current camera readout noise and pixel binning (Appendix~\ref{Appendix:image_process}), optimizing the trade-off between discrimination infidelity and imaging loss [Fig.~\ref{fig:appendix_img_chrac}(b) and Fig.~\ref{fig:appendix_imgloss_chrac}(a)] by reducing the imaging time would enable a further 2.5-fold reduction in imaging loss, to $\sim$\num{1.5e-4}.
Imaging losses as low as \num{3e-5} per image could be achieved using a CMOS camera with less than 1 electron readout noise and similar quantum efficiency.

\subsubsection{Background Gas Collision Loss}\label{sec:vacuum_loss}
With a detailed understanding of the losses associated with the imaging and cooling processes, we are able to extract accurate information about the background-gas-collision-limited lifetime.
In the lifetime sequence, we choose to lower the trap depth to $U_0/k_B=$ \SI{0.46}{\mK} [top panel of Fig.~\ref{fig:loss_charac}(c)], to mitigate residual parametric heating from intensity noise and intermodulation.
We experimentally verify that applying a \SI{25}{\ms} cooling pulse every \SI{10}{\second} results in an optimal sequence that balances the competing heating-induced losses and cooling losses.
To expedite the lifetime measurements, we increase the intensity of the repumping beams to saturate the transition (Appendix~\ref{Appendix:image_loss}). 
This adjustment enables us to detect approximately $140$ photons per cooling pulse, enough to determine atoms' survival.
Using MLE with censoring, we extract a lifetime of $2800^{+500}_{-400}$ \SI{}{\second} [Fig.~\ref{fig:loss_charac}(c)].
Note that this measurement was performed with the 2D MOT gate valve open, yet no background loading was detected in empty tweezer traps.
We further measured the trap lifetime with the gate valve closed, observing no noticeable difference (in contrast to Ref.~\cite{schymik2022scaling}) suggesting that the current lifetime limitation is not due to leakage from the 2D MOT.
To account for cooling losses, we applied a conservative correction factor of \num{2e-4} loss per cooling pulse, estimated based on the scattered photon number (Appendix~\ref{Appendix:image_loss}), to correct the raw lifetime result. 
This yields a cooling-loss-corrected atom trap lifetime of $3000^{+600}_{-500}$ \SI{}{\second}.
Using this lifetime, we estimate the equivalent equilibrium pressure of hydrogen at the atom array to be \SI{7e-13}{\mbar}.
This estimate accounts for the collisional loss cross section between $^{87}$Rb atoms and hydrogen molecules~\cite{arpornthip2012vacuum}, as well as the measured cold box temperature of \SI{45}{\K}, which affects the background gas collision loss rate by influencing gas number density and loss cross section~\cite{arpornthip2012vacuum,barker2022precise}.

The negligible line of sight for room-temperature background gas to the center of the cold box and the measured long atom trap lifetime compared to other room-temperature systems suggest that the pressure at the atoms is effectively decoupled from the external vacuum pressure during cryopumping operation.
The ultimate pressure is primarily limited by residual outgassing within the cold box at cryogenic temperatures.
Additionally, the use of low-outgassing materials and the adequate conductance to the exterior vacuum without cryopumping ensure favorable vacuum pressures for room-temperature operation, and enables lifetimes of up to \SI{30}{\second} at room temperature [Fig.~\ref{fig:loss_charac}(c)]. 

The atom trap lifetime achieved in our system is within a factor of two of the result described in Ref.~\cite{Schymik2021single}, which used a fully-enclosed \SI{4}{\K} cold shield. 
Ref.~\cite{manetsch2024tweezer} employed aggressive titanium sublimation pumping and multiple rounds of baking to achieve an atom trap lifetime of \SI{1400}{\second} at room temperature. 
Our system, without the need for a fully baked chamber, and with electrodes, plastic-insulated wires, coaxial cables, and other related components as described inside the cold box, demonstrates similar optical access with a twofold greater lifetime. 
The vacuum pressure could be further reduced in the future by extending the suspended 4 K finger into a pronged structure with increased surface area closer to the atoms while maintaining the current optical access.

\subsection{System Performance for Defect-Free Atom Array Assembly}
\begin{figure}[t]
    \centering
    \includegraphics[]{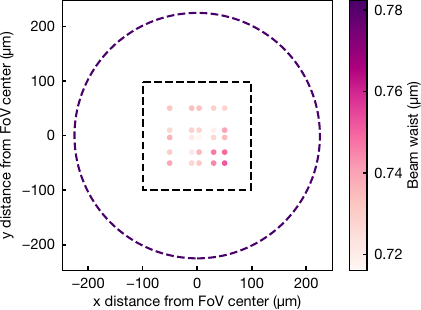}
    \caption{Performance of the optical tweezer system.
    Colored dots indicate tweezer beam waists measured within a \SI{100}{\um} square centered in the FoV in the focal plane of the objective.
    The waists are computed from the measured trap depth and trap frequency at various positions. 
    These measurements reveal a peak-to-peak waist variation of less than \SI{5}{\percent} across the atom array, with a mean waist value of \SI{0.72}{\um}, which corresponds to an average NA of 0.49.
    The dashed square outlines the region accessible by the AOD, with boundaries set by the limits of its deflection angle.
    The dashed circle demarcates the expected FoV of the objective, with a diameter of \SI{450}{\um}, based on the optical design. 
    The color of the dashed circle represents the expected beam waist of \SI{0.78}{\um} at this diameter based on objective specification, which is extrapolated from the largest measured waist of \SI{0.75}{\um} and rescaled using a separate measurement of the objective's NA with a pinhole.
    }
    \label{fig:beam_waist}
\end{figure} 

\begin{figure*}[th]
    \includegraphics[]{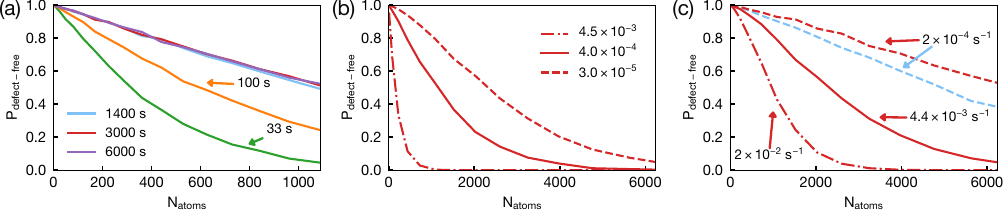}
    \caption{
    Simulation results illustrating the probability of assembling a defect-free array as a function of the number of atoms in the array, considering various combinations of losses. 
    (a) Dependence on atom trap lifetime given our measured imaging loss and cooling loss. Trace colors represent different lifetimes: green (\SI{33}{\second}), orange (\SI{100}{\second}), blue (\SI{1400}{\second}), red (\SI{3000}{\second}) and purple (\SI{6000}{\second}).
    (b) Dependence on imaging loss per cycle given our measured atom trap lifetime and cooling loss. Trace styles indicate different imaging losses per pulse: solid (\SI{4e-4}, our measured value), dashed (\SI{3e-5}), dash-dot (\SI{4.5e-3}). 
    (c) Dependence on cooling loss given improved hypothetical imaging loss of \SI{3e-5} with our measured (red) and a \SI{1400}{\second} (blue) atom trap lifetime.
    Trace styles denote different cooling loss rate: solid (\SI{4.4e-3}{\per\second}, our measured value), dashed (\SI{2e-4}{\per\second}, for both red and blue traces), dash-dot (\SI{2e-2}{\per\second}).
    }
    \label{fig:rearrangement}
\end{figure*} 
Having demonstrated long atom trap lifetimes and low cooling and imaging losses, we now explore their implications for assembling large defect-free atom arrays\textemdash an increasingly important resource for neutral atom quantum simulation and quantum computing\textemdash in our cryogenic setup. 
Such arrays rely on high-performance lenses to create many optical tweezers with homogeneous traps. 
Our current experimental setup uses a \SI{100}{\um} square FoV, within which beam waists show less than \SI{5}{\percent} variation (Fig.~\ref{fig:beam_waist}), inside the AOD-accessible region.
The objective is designed for a circular FoV of \SI{450}{\um} diameter and the measurements over a \SI{100}{\um} FoV are consistent with this specification. 
In comparison, previous work in a long-trap-lifetime cryogenic apparatus~\cite{Schymik2021single} achieved a square FoV of \SI{50}{\um}. 

The ability to create defect-free arrays is analyzed theoretically through Monte Carlo simulations.
We initiate the simulations with a \SI{90}{\percent}-filled array supplemented by a reservoir, which could be achieved with a pre-rearranged sample or deterministic loading. 
We report the final steady-state filling fraction, which should be independent of this initial condition.
In our simulations, we employ a square array geometry with parallel movement within a single row or column via crossed AODs. 
Specifically, we use unidirectional row compression~\cite{endres2016atom} to fill vacant sites with extra atoms from the reservoir, which requires moving all atoms on one side of the vacant site along the grid. 

Under these assumptions, it can be shown that there exists a threshold for losses caused by atom movement operations that scales as $1/\sqrt{N}$, where $N$ is the total atom number.
Above this threshold, rearrangement rounds only decrease the filling fraction, whereas below the threshold additional rearrangement cycles increase the filling fraction up to a saturation level defined by the losses from other channels discussed below. 
Consequently, a typical atom movement loss of less than \SI{1}{\percent}~\cite{barredo2016atom} is not relevant for assembling an array of a few thousand atoms.
From our simulations, we observe that the filling fraction reaches a steady state after 5 rounds of rearrangement.  
Therefore, a 5-round rearrangement cycle is used throughout the simulations.

The simulation incorporates background gas collision loss and assumes an imaging time of \SI{14}{\ms}, \SI{1}{\ms} rearrangement per row, and \SI{20}{\ms} idle time for data transfer and calculation per rearrangement cycle~\cite{bluvstein2024logical, wang2023accelerating}. 
The simulation also accounts for losses due to imaging and cooling, where images are taken before and after rearrangement, with cooling light applied throughout the atom movement.
Unless otherwise specified, the simulation assumes default values for imaging loss of \SI{4e-4} per pulse and a cooling loss rate of \SI{4.4e-3}{\per\second}.

We investigate the dependence of the defect-free probability for typical values of atom trap lifetime obtained in various optical tweezer experiments (Fig.~\ref{fig:rearrangement}(a)).
We consider lifetimes of \SI{33}{\second}~\cite{tian2023parallel}, \SI{100}{\second}, \SI{1400}{\second}~\cite{manetsch2024tweezer} (longest reported at room temperature) and \SI{6000}{\second}~\cite{Schymik2021single} (best at cryogenic temperature), and compare these with our result of \SI{3000}{\second}.
With our measured imaging loss, lifetimes above \SI{1000}{\second} show notable enhancements in the probability of generating large defect-free arrays compared to shorter atom trap lifetimes.
The marginal variation observed across lifetimes exceeding $\sim\SI{1000}{\second}$ suggests that the limit to high defect-free probability is not the atom trap lifetime given our imaging and cooling loss rates.

To further explore this limit, we compare our measured imaging loss with a considerably improved hypothetical value of \SI{3e-5} (Sec.~\ref{sec:cooling_and_imaging_loss}) assuming our measured atom trap lifetime [Fig.~\ref{fig:rearrangement}(b)].
This value, which is an order of magnitude smaller than the current best reported imaging loss \cite{manetsch2024tweezer}, represents a feasible target for future investigations.
Achieving this could potentially involve enhancing collection efficiency through optical cavities~\cite{shadmany2025cavity, hu2025site} or reducing sensor readout noise, both of which could decrease the number of scattered photons needed for discrimination and thus reduce imaging loss (Appendix~\ref{Appendix:image_loss}).
Such low imaging loss would enable a nearly \SI{90}{\percent} defect-free probability with up to 1000 atoms, fully leveraging the cryogenic advantages.

We also explore the dependence on cooling loss assuming our measured atom trap lifetime and an imaging loss of \num{3e-5} [Fig.~\ref{fig:rearrangement}(c)].
We consider cooling loss rates of \SI{2e-4}{\per\second}, \SI{4.4e-3}{\per\second}, and \SI{2e-2}{\per\second}. 
This analysis demonstrates that, with \SI{2e-4}{\per\second} cooling loss rate, the defect-free probability for approximately 1000 atoms can be further enhanced to greater than \SI{95}{\percent} and the system size can be extended to a few thousand atoms, while still maintaining a practical duty cycle for achieving a defect-free array.

The simulation results highlight the need for long atom trap lifetimes on the order of thousands of seconds, while reducing imaging and cooling losses, to realize defect-free atom arrays comprising thousands of atoms\textemdash a crucial step toward scalable quantum computation and simulation based on neutral atoms. 
In particular an atom trap lifetime of \SI{3000}{\second} becomes advantageous compared to \SI{1000}{\second} when the cooling loss rate approaches \SI{1e-4}{\per\second} and the imaging loss per pulse is near \num{1e-5}, both corresponding to a roughly tenfold improvement compared to our measured best values [red and green dashed lines in Fig.\ref{fig:rearrangement}(c)].

\section{Single Qubit Microwave Rotation} \label{Microwave_qubit}

\begin{figure*}[th]
    \includegraphics[]{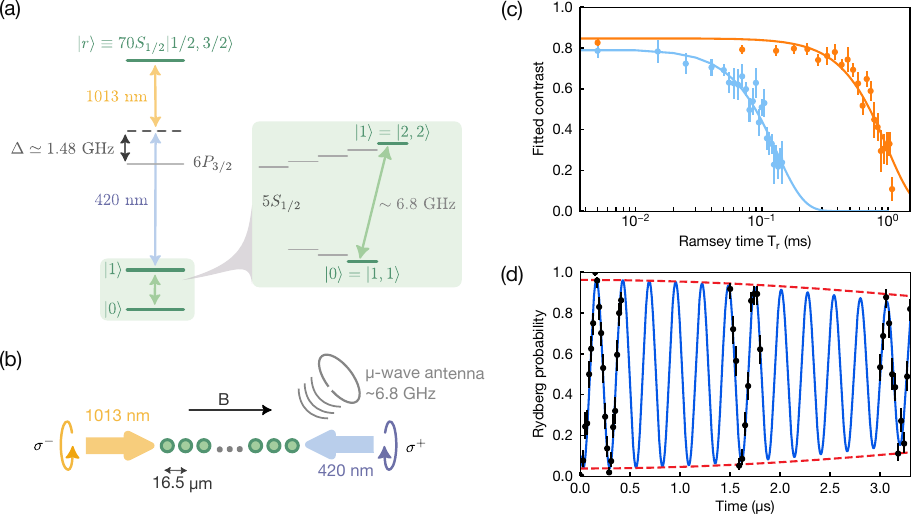}
    \caption{
    Single-atom control of ground-state and Rydberg qubits.
    (a) Level structure of $^{87}$Rb with the relevant qubit and Rydberg states. 
    The ground-state qubit is encoded in the $\ket{1}\equiv\ket{F=2,m_F=2}$ and $\ket{0}\equiv\ket{F=1,m_F=1}$ states of the 5S$_{1/2}$ ground hyperfine manifolds. 
    The Rydberg excitation, between $\ket{1}$ and $\ket{r}\equiv 70 S_{1/2} \ket{m_s=1/2, m_I=3/2}$, is a two-photon process with a detuning of $\Delta \simeq \SI{1.48}{\GHz}$ from the intermediate 6P$_{3/2}$ state. 
    (b) Qubit addressing geometry. 
    Ground-state qubit control is achieved with a microwave antenna globally addressing all atoms with a Rabi frequency of $2\pi \times$\SI{5.6}{\kHz}. 
    Rydberg excitations are driven by two circularly polarized counter-propagating beams at \SI{420}{\nm} and \SI{1013}{\nm}.
    The microwave antenna is also used for Rydberg state discrimination.
    (c) Ramsey contrast decay as a function of the time $T_r$ between two $\pi/2$ pulses on the ground-state qubit. 
    Blue (orange) points represents Ramsey sequence without (with) a spin-echo pulse.
    Each data point represents the fitted contrast when scanning the phase of the second $\pi/2$ pulse. 
    The $T_2^\star$ coherence time of \SI{131(5)}{\us}\textemdash fitted with a Gaussian profile\textemdash corresponds to a magnetic field RMS fluctuation of \SI{0.82(3)}{\milli G} over 10 minutes (measurement time for each data point). The spin-echoed $T_2$ is \SI{0.95(5)}{\ms}.
    (d) Ground-Rydberg Rabi oscillations.
    The blue line is the fitted decaying cosine function with Gaussian envelope, with a Rabi frequency of $2\pi \times$\SI{3.8}{\MHz} and Gaussian decay $1/e$ time constant of \SI{5.4(7)}{\us}.
    The red dashed line is the calculated Gaussian envelope (\SI{7.6(1)}{\us} $1/e$ time constant) based on independent estimates of the relevant decoherence channels in our system given current laser system performance (Appendix~\ref{Appendix:decoherence_channel}).
    Each data point is derived from an average of 50 repeated measurements.
    }
    \label{fig:qubit_control}
\end{figure*} 

The qubit states are encoded in the $5S_{1/2}$ hyperfine manifold as $\ket{F=2, m_F=2} \equiv \ket{1}$ and $\ket{F=1, m_F=1} \equiv \ket{0}$, chosen for their simplicity in state preparation and magnetic field sensitivity, making them well-suited for characterizing the magnetic field environment in our cryogenic setup.
After loading atoms into the optical tweezers, a \SI{3.5}{G} quantization field is applied in the radial plane of the trapping potential, in the $-x$ direction (Fig.~\ref{fig:setup_overview}).
Atoms are initialized in $\ket{1}$ via optical pumping through the $F=2 \leftrightarrow F^{\prime}=2$ $D_2$ transition.
Readout of the qubit states is realized by driving the cycling transition of $\ket{1} \leftrightarrow 5P_{3/2}\,\ket{F^{\prime}=3, m_{F^{\prime}}=3}$ using the same beam for optical pumping.
This heats atoms in state $\ket{1}$, leading to atom loss, and a subsequent RPGC imaging pulse reveals the atoms remaining in state $\ket{0}$.

Global qubit rotations are achieved by driving microwave transitions using the half-wave antenna positioned inside the cold box (Fig.~\ref{fig:setup_overview}).
The tweezer trap depth is adjusted to \SI{84}{\uK} for ground state qubit rotations, mitigating the impact of differential light shifts from the trapping light.
The achievable Rabi frequencies for driving hyperfine transitions, projected onto the quantization axis, is characterized as $2\pi\times$(\SI{2}{\kHz}, \SI{23}{\kHz} and \SI{6}{\kHz}) for ($\sigma^-$, $\pi$ and $\sigma^+$) polarization components respectively, for state $\ket{0}$.
Using microwave transitions, we can measure the magnetic field dynamics during field switching, caused by eddy currents in the cold box. 
Along the vertical ($z$) axis, the measured eddy current decay time constant is \SI{5(0.6)}{\ms}, considerably shorter than that measured along the $x$ and $y$ axes, respectively \SI{106(8)}{\ms} and \SI{63(8)}{\ms}, due to the slit in the cold box that breaks the current loop around the $z$ axis (Fig.~\ref{fig:setup_overview}).
The eddy current time constants are obtained through least-squares fits.
In practice, we allow \SI{700}{\ms} for the bias field to stabilize before performing state preparation.
To mitigate the effects of eddy currents, additional slits could be introduced along the $x$ and $y$ axes.
Alternatively, a separate room-temperature UHV chamber could be used for the 3D MOT, from which the atoms would be transferred to the cryogenic science chamber, which maintains a constant static magnetic field.

The coherence of the ground state qubit is evaluated using a Ramsey sequence [Fig.~\ref{fig:qubit_control}(c)].
Each data point corresponds to the contrast of the sinusoidal oscillations obtained by scanning the phase of the second $\pi/2$ pulse.
The $2\pi\times$\SI{6}{\kHz} Rabi frequency, in conjunction with shot-to-shot magnetic field fluctuations and variations in differential light shifts due to trap depth disorder, limits the initial Ramsey contrast to 0.8.
The contrast decay is modeled with a $e^{-t^2/\tau^2}$ Gaussian profile, with the amplitude included as a fitted parameter. This fit yields a pure dephasing time $T_2^\star$ of \SI{131(5)}{\us} and $T_2$ of \SI{0.95(5)}{\ms}, with least-squares fits.
For $T_2^\star$, the Gaussian decay constant $\tau$ is related to the shot-to-shot transition frequency fluctuation $\Delta_0$ by $\tau=\sqrt{2}/\Delta_0$. 
Consequently, the RMS magnetic field fluctuation is calculated to be \SI{0.82(3)}{\mG} over the 10 minute measurement time window used for each data point acquisition.
This sub-mG scale fluctuation, primarily driven by noise from the laboratory ac power mains, is typical of fluctuations observed in atomic physics laboratories, highlighting the stability of the magnetic environment in our cryogenic setup.

\section{Rydberg Excitation} 
\label{sec:Ryberg_excitation}
Rydberg control is achieved via a two-photon excitation of \SI{420}{\nm} and \SI{1013}{\nm} through the intermediate $6P_{3/2}$ state in the inverted scheme~\cite{bernien2017probing} with \SI{1.48}{\GHz} detuning [Fig.~\ref{fig:qubit_control}(a)]. 
The final Rydberg state is denoted as $\ket{r}\equiv70S_{1/2}\ket{m_s=1/2, m_I=3/2}$ where $m_s$ ($m_I$) represents the magnetic quantum number for electron (nuclear) spin.
The \SI{420}{\nm} light is produced through second harmonic generation of the fundamental light from a Ti:Sapphire laser.
The \SI{1013}{\nm} light is amplified using a fiber amplifier, with the seed provided by the amplified cavity-filtered transmission~\cite{levine2018high} from a diode laser and a semiconductor optical amplifier.
Both the Ti:Sapphire and diode laser are frequency stabilized to a cavity made with an ultra-low-expansion glass spacer.
The \SI{420}{\nm} (\SI{1013}{\nm}) beam is directed through the side cold window, as depicted in Fig.~\ref{fig:setup_overview}, counter- (co-) propagating with respect to the quantization axis with $\sigma^+$ ($\sigma^-$) polarization.
The waists are approximately \SI{40}{\um} for  both the \SI{420}{\nm} and \SI{1013}{\nm} beams.
The corresponding single-photon Rabi frequencies are $2\pi\times\SI{145}{\MHz}$ and $2\pi\times\SI{78}{\MHz}$, respectively.

For Rydberg-related experiments, the tweezers are configured in a single row comprising 13 traps, spaced by \SI{16.5}{\um} to suppress the van der Waals interaction strength to well below the ground-Rydberg Rabi frequency.
During the MOT loading stage, UV LEDs at \SI{365}{\nm} are utilized to reduce patch charges on the dielectric surfaces~\cite{bernien2017probing} within the cold box.
Furthermore, electrodes (Fig.~\ref{fig:setup_overview}) are used to cancel any residual electric field. 
While the applied field strengths differ between room-temperature and cryogenic operation, in both cases the overall strengths remain below \SI[per-mode = symbol]{0.4}{\volt\per\cm}.

To prepare for Rydberg excitation, the atoms are initially optically pumped into state $\ket{1}$. 
The trap depth is then adiabatically lowered to \SI{84}{\uK} before excitation to minimize temperature-induced dephasing.
During the excitation, the optical tweezers are turned off to avoid losses from anti-trapping of the Rydberg state.
After excitation, the traps are rapidly turned on to approximately \SI{850}{\uK} within \SI{2}{\us}, and a \SI{6.35}{\GHz} microwave pulse is applied for \SI{100}{\us} using the same antenna and electronics described in Sec.~\ref{Microwave_qubit}.
This microwave pulse causes loss of the atoms in the Rydberg state~\cite{graham2019rydberg}.
Subsequent RPGC imaging reveals the atoms in the ground state, with the absence of atom signals indicating the presence of atoms in the Rydberg state, up to state preparation and detection errors.

The resulting Rabi oscillations between the ground state $\ket{1}$ and the Rydberg state $\ket{r}$ are shown in Fig.~\ref{fig:qubit_control}(d).
Each data point is based on an average of 50 repeated measurements, and the total duration of the experiment is around 10 minutes.
A Gaussian decaying cosine function, with an envelope described by $e^{-t^2/\tau^2}$, is fitted to the data, yielding a Rabi frequency of $2\pi \times$\SI{3.8}{\MHz} and a decay constant $\tau$ of \SI{5.4(7)}{\us}.
The first three data points are excluded from the fit to account for the finite rise and fall time of the acousto-optic modulator controlling the beam intensity.
We also performed the Rabi experiment with the cold box at room temperature, resulting in a decay time of \SI{6(1)}{\us}.
To characterize sources of decoherence, we performed a Monte Carlo simulation of the Rydberg Rabi oscillations, accounting for several decoherence channels that were independently measured and calibrated (Appendix~\ref{Appendix:decoherence_channel}).
These include laser phase and intensity noise, finite intermediate and Rydberg state lifetime, finite atomic temperature, beam pointing fluctuations, pulse area variations, and site-dependent inhomogeneities in both the Rabi coupling strengths and resonance frequencies.
We find that technical considerations dominate the decay rate.
The simulation yields a fitted Gaussian decay time constant of \SI{7.6(1)}{\us} [red dashed line in Fig.~\ref{fig:qubit_control}(d)], near to though slightly exceeding the experimental data.

While a comprehensive understanding and improvement of Rydberg coherence will be addressed in future work, the simulation indicates that the primary limitation on coherence time is currently due to shot-to-shot optical pulse area fluctuations.
The near agreement between the experimental results and the Monte Carlo simulation (Appendix~\ref{Appendix:decoherence_channel}) indicates that the cryogenic setup does not introduce significant additional decoherence compared to room temperature operation for our current coherence time.

\section{Outlook}
In this work, we present the successful integration of cryogenic techniques with advanced control of neutral atoms.
Our setup features low wavefront distortion, large optical access, stable optical alignment and minimal vibrations.
We have characterized the loss channels in the trapping system and achieved a single-atom trapping lifetime of \SI{3000}{\second} through cryopumping of background gas.
The extended lifetime allows us to measure imaging and cooling losses with high accuracy, offering deeper insights into these loss mechanisms~\cite{martinez2018state} and paving the way for further improvements. 
Further, this extension of atom trap lifetime can reduce the overhead for detecting and correcting qubit loss errors~\cite{stricker2020experimental, chow2024circuit}.
Based on our technical advancements, we also discuss the implications of assembling a large, defect-free array of such long-lived atomic qubits. 
Additionally, we demonstrate coherent control of both ground and Rydberg state qubits, validating the technical feasibility of implementing high-fidelity single-qubit operations and precise Rydberg state control within a cryogenic environment.

Our platform also opens new possibilities for controlling the BBR environment in Rydberg atom experiments using optical tweezers.
With the addition of an ITO coating to the cold window to shield against room-temperature microwave-frequency BBR, the system is well-suited to extend Rydberg lifetimes at both single-particle and many-body scales.
Achieving a well-controlled cryogenic microwave environment will support the advance of high-fidelity quantum gates~\cite{evered2023high} and enable the study of complex quantum many-body dynamics in Rydberg systems over extended timescales~\cite{boulier2017spontaneous, zeiher2017realization}. 
This includes investigations into non-equilibrium phenomena like squeezing and equilibrium phases such as frustrated ground states~\cite{eckner2023realizing, semeghini2021probing}.

Beyond its immediate implications for Rydberg physics, our platform is applicable to a broad range of neutral atom species and molecules.
This flexibility unlocks new possibilities for scalable, high-fidelity quantum computing, quantum simulations of large-scale many-body systems, and enhanced metrological stability.

\section{Acknowledgements} This material is based upon work supported by the U.S. Department of Energy, Office of Science, National Quantum Information Science Research Centers, Quantum Systems Accelerator. Additional support is acknowledged from NSF 2210527, NSF QLCI award OMA – 2016244, ARO grant W911NF-19-1-0223, the Swiss National Science Foundation under grant 211072 (MM), Atom Computing, NIST, and the Baur-SPIE Chair at JILA.  We thank Tobias Thiele, Mark Brown, Max Kolanz, Prithvi Raj Datla, and Jamie Boyd for helpful discussions and assistance. We also thank Kyungtae Kim and Yue Jiang for a careful reading of this manuscript.

\appendix
\section{Cryocooler}\label{Appendix:Cryocooler}
The closed cycle cryocooler from ColdEdge technologies consists of two helium loops as discussed in Sec.~\ref{cryostat_and_vacuum_chamber}.
The first loop drives a Sumitomo RDK-415D2 GM cryocooler, providing up to \SI{1.5}{\watt} of cooling power at \SI{4}{\K}. 
The second loop is driven by Sumitomo HC-4E1 or KNF N630.15 recirculating compressor and passes through a control manifold before entering the flexible transfer line. 
The control manifold includes valves to isolate the transfer line from the recirculating compressor, a pressure regulator to control the helium flow, a vacuum port for evacuating the loop, and a charging port for helium recharge. 
This secondary loop interfaces with the GM cryocooler located under the optical table through heat exchangers anchored to the cold stages of the GM cryocooler. 
The helium cools to approximately \SI{6}{\K} before exiting via a flexible coaxial helium transfer line.
At the end of the transfer line, a capillary restriction further reduces the helium gas temperature to below \SI{4}{\K} through the Joule-Thomson effect.
The flexible transfer line is then connected to the UHV cryostat using a locking collar and o-ring seals, and is removable without breaking vacuum in the experimental system [Fig.~\ref{fig:mechanical_detail}(b)].

\section{Cryopumping Surface Regeneration}\label{Appendix:regeneration}
The cold finger surface gradually saturates with cryopumped gases, increasing the background pressure in the cold box and reducing the atom trap lifetime proportionally, with an exponential time constant of \SI{17}{\hour}.
The \SI{4}{K} cold finger’s low thermal mass and thermal isolation enable daily regeneration of the cryopumping surface via integrated heaters to restore the base pressure and atom trap lifetime, an approach that would be difficult without a thermal design compatible with in-situ heating~\cite{pichard2024rearrangement}.
During regeneration, \SI{5}{\watt} of power is applied for a few hours, raising the cold finger temperature to \SI{22}{K} while maintaining the cold box temperature below \SI{70}{K}.
This causes a pressure spike as measured by the ion pump current, peaking at $\sim$\SI{1e-9}{\mbar}, due to hydrogen release from \SI{4}{K} components.
Once heating stops, the cold finger cools to \SI{4}{K} within seconds, and the ion pump pressure drops below \SI{1e-11}{\mbar}, confirming efficient cryopumping of hydrogen.

\section{Image Processing and Discrimination Fidelity}\label{Appendix:image_process}

\begin{figure*}[htbp]
    \includegraphics[]{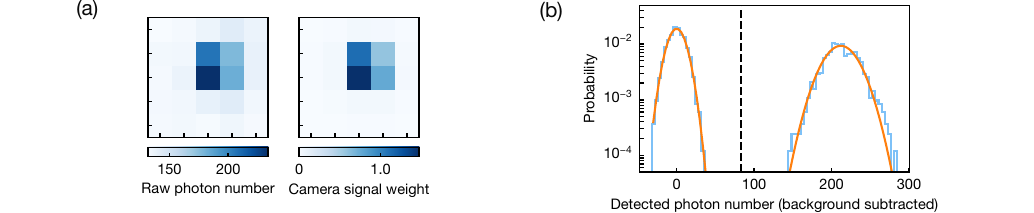}
    \caption{
    Image processing procedure and atom discrimination.
    (a) Left panel: an averaged image of atoms from a specific trap site. Right panel: the weight function, derived from fitting the averaged atom image on the left with a coarse-grained PSF, used to determine the detected photon number for this trap site.
    (b) Extracted photon number histogram for a single trap (blue) and Gaussian-fitted photon number statistics for both the atom signal and background (orange). The dashed line represents the optimal threshold that discriminates between the two Gaussian distributions, achieving a discrimination infidelity of less than \num{1e-6}.
    }
    \label{fig:appendix_img_chrac}
\end{figure*} 

Photons scattered during the imaging process are collected by the \SI{0.55}{NA} objective and imaged onto a scientific CMOS camera (Andor Marana 4.2B-11). 
This camera has a quantum efficiency of \SI{65}{\percent} at \SI{780}{\nm} and a gain of $1.41$ (converting from camera signal to detected photon number without correcting for quantum efficiency).
The average raw atom image of one particular trap site is shown in Fig.~\ref{fig:appendix_img_chrac}(a) with $2 \times 2$ pixel binning and pixel values offset by a constant photon number introduced by the camera firmware.

To determine the total number of detected photons for each trap site, we use a weight function obtained by fitting a point spread function (PSF) with an overall offset to the average atom images for each site.
The averaged atom images are cropped around their center to a $5\times5$ region, with the photon number in each pixel denoted as $C_{ij}$, corresponding to the pixel in row $i$ and column $j$.
The PSF on the camera sensor plane is modeled as an Airy disk with the functional form
\begin{equation}
    A\left[\frac{J_1\left(M\sqrt{(x-x_0)^2 + (y-y_0)^2}\right)}{M\sqrt{(x-x_0)^2 + (y-y_0)^2}}\right]^2 + B,
\end{equation}
where $J_1$ is the Bessel function of the first kind of order one, $A$ represents the amplitude, $M$ is a scale factor determined by the imaging system, $x_0$ and $y_0$ are the coordinates of the PSF center, and $B$ accounts for the offset in the atom image.
To account for pixelation, the PSF is integrated over each pixel region, producing a binned representation that captures the total signal distribution on the discrete image grid, and used for fitting the averaged atom image to extract $A$, $M$, $x_0$, $y_0$ and $B$.
With the fitted parameters, the binned PSF is evaluated without the fitted offset $B$ and denoted as $F_{ij}$.
The weight function $W_{ij}$ [Fig.~\ref{fig:appendix_img_chrac}(a)] is then defined for each pixel as 
\begin{equation}
    W_{ij} = F_{ij}\frac{\Sigma_{i,j}F_{ij}}{\Sigma_{i,j}F_{ij}^2},
\end{equation}
where the sum is performed over the $5\times5$ region.
For subsequent atom imaging, the raw images are multiplied by this weight function $W_{ij}$ pixel by pixel and the total detected photon number for each trap site are obtained by summing all pixel values of the corresponding weighted images.
Relative to a uniform weight function $W_{ij}=1$, this choice of weight function maintains the total detected photon number for a fluorescing atom but reduces readout noise, albeit still above the single-binned-pixel level, thus lowering discrimination infidelity. 

Throughout this manuscript, the imaging scattering rate refers to the rate determined by the camera-collected fluorescence photons, rescaled by the calculated \SI{4}{\percent} collection and detection efficiency of the imaging system.
The histogram for a particular trap site is depicted in Fig.~\ref{fig:appendix_img_chrac}(b).
Note that the histogram's $x$-axis is shifted so that the background peak is at zero.
The bimodal distribution in the histogram is fitted with two Gaussians, and their parameters are used to determine the optimal threshold for atom discrimination, minimizing the probabilities of false positives and false negatives.
Applying the same procedure to all trap sites yields discrimination infidelities of less than \SI{1e-6}{} across all sites using the standard imaging parameters described in the main text.

\section{Imaging Loss and Scaling}\label{Appendix:image_loss}

\begin{figure*}[htbp]
    \includegraphics[]{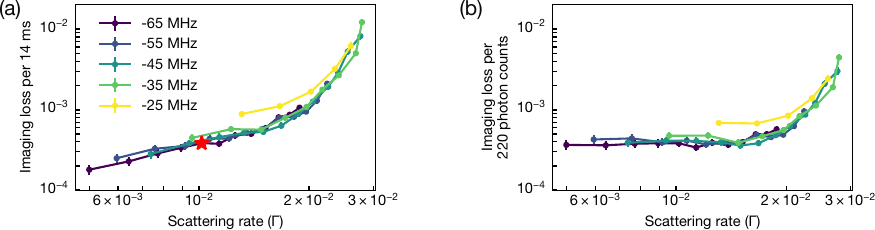}
    \caption{
    Imaging loss investigation in a larger parameter space. 
    (a) Measured loss per \SI{14}{\ms} imaging pulse as a function of scattering rate at various detunings. 
    The detuning values shown in the legend are relative to the free-space resonance. 
    The red star indicates the standard imaging parameters used in the main text.
    (b) Rescaled imaging loss for fixed discrimination infidelity of \SI{1e-6}{} and variable imaging duration, with total camera detected photons matching the values from the standard imaging setup used in the main text. 
    After rescaling, the losses flatten out and do not show the decreasing trend as the scattering rate reduces. 
    }
    \label{fig:appendix_imgloss_chrac}
\end{figure*} 
The standard imaging light is red-detuned by \SI{65}{\MHz} from the free space resonance. 
Considering the trap-induced differential light shift, the detuning to the in-trap resonance is approximately \SI{90}{\MHz}.
The scattering rate calculated from beam power and detuning is a factor of 4 larger than the one inferred from camera signals.
This discrepancy may arise from atomic motion and differential light shifts in the tweezer.

We investigate the dependence of imaging loss on detuning and beam power for a \SI{14}{\ms} imaging pulse [Fig.~\ref{fig:appendix_imgloss_chrac}(a)] using the procedure outlined in the main text.
These results demonstrate a clear trend of diminishing imaging loss per \SI{14}{\ms} image as the scattering rate decreases.
However, the resulting reduction in photon number per image can impact discrimination fidelity.
To account for this, Fig.~\ref{fig:appendix_imgloss_chrac}(b) plots the loss per image normalized to a fixed detected photon number of 220, corresponding to the average photon number with our standard imaging parameters [red star in Fig.~\ref{fig:appendix_imgloss_chrac}(a)]. 
This gives a fixed discrimination infidelity of \SI{1e-6}{} rather than a fixed imaging duration.
Following this rescaling, the imaging losses saturate at a constant value for scattering rates below $0.1\,\Gamma$.
The slightly higher imaging loss at \SI{-25}{\MHz} detuning is likely due to sub-optimal PGC performance with smaller detuning~\cite{dalibard1989laser}.

This observation suggests a limit to RPGC imaging loss, which depends simply on the number of scattered photons rather than on the imaging time or scattering rate individually, as long as the scattering rate is well below the transition linewidth.
A hypothesis for the physical reason behind this dependence for our atom and trap wavelength is the anti-trapping of the excited $5P_{3/2}$ state~\cite{martinez2018state}. 
One possible way to reduce the effect of anti-trapping on imaging loss would be to use stroboscopic imaging~\cite{hutzler2017eliminating}, or reduce the number of scattered photons per image by increasing the collection efficiency and lowering the camera noise, thus keeping the same discrimination fidelity.

It would be interesting to explore whether other imaging schemes, such as LGM~\cite{blodgett2023imaging} exhibit the same behavior.
Additionally, examining this phenomenon in other species that can support trapped excited states~\cite{manetsch2024tweezer}, and in alkaline earth species using narrow-line Sisyphus cooling~\cite{covey20192000}, would provide further insights.

\section{Decoherence Channels in Rydberg Excitation}\label{Appendix:decoherence_channel}
\begin{figure}[htbp]
    \centering
    \includegraphics[]{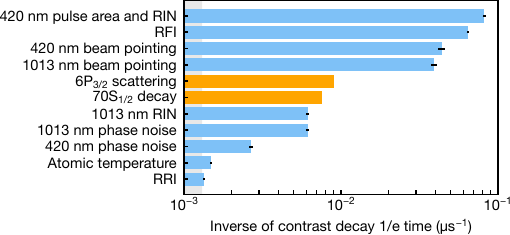}
    \caption{Decoherence channels and their contributions to the Rydberg Rabi oscillation contrast decay in this work. 
    Orange (blue) bars represent channels modeled with an exponential (Gaussian) decay, as detailed in the main text.
    The gray shaded region indicates the level of numerical error in the simulation.
    RFI denotes Rabi frequency inhomogeneity, and RRI represents Rydberg resonance inhomogeneity, both indicating site-dependent variations in the system.
    The Rydberg lifetime is conservatively set to \SI{100}{\us}, smaller than the room-temperature value of \SI{147}{\us}.
    Simulation parameters are based on the those used for the Rabi oscillation in Sec.~\ref{sec:Ryberg_excitation}.
    }
    \label{fig:appendix_rydberg_simulation}
\end{figure} 

Here we provide a summary of the decoherence channels in the Rydberg Rabi-oscillation data and their calculated decoherence rates. Among the decoherence mechanisms, Rydberg state decay and scattering from the intermediate $6P_{3/2}$ state are modeled as jump operators in the Lindblad master equation.
The calculated $70S_{1/2}$ Rydberg state lifetime is \SI{147}{\us} at \SI{300}{\K} and \SI{306}{\us} at \SI{45}{\K}.
The experimental measurement of the Rydberg lifetime at cryogenic temperatures will be the subject of future investigation, and we conservatively set the cryogenic lifetime to be \SI{100}{\us} in the simulation. 
The effect of scattering from the intermediate $6P_{3/2}$ state is calculated using the single-photon Rabi frequencies and intermediate state detuning values given in Sec.~\ref{sec:Ryberg_excitation}.
Exponential fits to the numerically solved master equations provide the $1/e$ time constants for contrast decay (orange bars in Fig.~\ref{fig:appendix_rydberg_simulation}).

Additional decoherence channels include laser phase noise and relative intensity noise (RIN), beam pointing fluctuations, atomic motion and positional spread due to finite temperature, pulse area variations, and site-dependent inhomogeneities in both the Rabi coupling strength and resonance frequency.
For these decoherence channels, we perform Monte Carlo simulations by numerically solving the Schr\"{o}dinger equation with empirically derived parameter variations or noise as discussed below. 
The simulation for each channel is conducted with an otherwise ideal Hamiltonian over 1000 iterations and the averaged result is fitted to a decaying cosine with a Gaussian envelope to extract the corresponding $1/e$ time constants and bootstrapped uncertainties (blue bars in Fig.~\ref{fig:appendix_rydberg_simulation}).
The precision of the numerical solver is evaluated by disabling all decoherence channels; the resulting fitted decoherence time constant (gray shaded area in Fig.~\ref{fig:appendix_rydberg_simulation}) characterizes the effect of finite numerical precision in the simulations.

The spectra of the laser phase noise and RIN are measured individually for each beam using a phase noise analyzer (OEwave OE4000) and a fast photodetector, respectively.
Based on these spectra, random time series of phase and intensity fluctuations are generated over a frequency range defined by the inverse of the total evolution time (\SI{5}{\us}) and the simulation time step (\SI{1}{\ns}).
Time series of laser phase are directly incorporated into the Schr\"{o}dinger equation to account for its effect on the Rabi oscillation dynamics, while time series of intensity fluctuations are combined with the shot-to-shot fluctuation from the RIN spectrum or pulse area fluctuation as discussed below before incorporation.

In the Rydberg sequence, the \SI{1013}{\nm} beam is turned on before the excitation, and the \SI{420}{\nm} beam is pulsed to gate the excitation time. 
Shot-to-shot intensity fluctuations of the \SI{1013}{\nm} beam due to RIN are modeled by integrating its intensity noise spectrum from \SI{20}{\hertz} to \SI{200}{\kHz}, the lower bound of the sampled simulation frequency range,  and treating the noise as quasi-static within each simulated experimental shot.
This approximation is valid because experimental runs are separated by intervals on the order of seconds.
For the \SI{420}{\nm} beam, pulse area fluctuations contribute to shot-to-shot intensity and the resulting distribution is used to sample the overall Rabi frequency for each shot.
Time series generated from the RIN spectra of both beams are combined with their respective shot-to-shot intensity fluctuations for the Monte Carlo simulation.

Additionally, beam pointing fluctuations contribute to shot-to-shot intensity fluctuations.
Beam positions are recorded during the experiment using beam pickoffs and cameras, and the distribution of beam center coordinates is used to generate static Rabi frequencies for each shot based on the measured beam waist.

Intensity variations of either beam at the atoms, which arise from a number of different channels, also cause changes in the light shift on the Rydberg transition.
This relationship is calibrated for each beam by Ramsey spectroscopy and is used in the simulation.

For the finite temperature effect, the spatial and velocity distributions of the atoms are considered. 
The atomic temperature right before applying Rydberg excitation is measured to be \SI{5}{\uK}.
The corresponding velocity spread gives rise to a Gaussian distribution of Doppler-shifted detunings.
Based on the measured trap frequency, the atomic spatial spread is calculated. 
The axial spread, combined with the measured Rydberg beam waist, contributes to shot-to-shot variations in the Rydberg resonance and Rabi coupling strength.

Static inhomogeneities in Rabi coupling strength across tweezer sites are characterized through high-statistics Rabi oscillation experiments, where the oscillation frequency at each site is analyzed, and the standard deviation across all sites is used as input for the simulations.
Similarly, site-dependent inhomogeneities in the transition frequencies were determined by measuring the phase accumulation rate in Ramsey spectroscopy, with both the \SI{1013}{\nm} and \SI{420}{\nm} beams turned off during the wait time.

The loss resulting from the turning off of the traps during Rydberg excitation is estimated to be less than \SI{1}{\percent} based on the measured temperature and is consistent with experimental observations. 
Due to its minimal contribution, this loss is neglected in the analysis.

Taking into account all the decoherence channels listed above, we obtain an effective overall Gaussian decay profile with a $1/e$ time constant for contrast decay of \SI{7.6(1)}{\us}.
The decay is primarily caused by pulse area fluctuations of the \SI{420}{\nm} beam, with an average variation around \SI{1}{\percent} over different pulse durations, and by inhomogeneities in the Rabi frequency across sites, which exhibit a standard deviation of \SI{0.4}{\percent}.

Additionally, during routine operation, we observed fluctuations in Rydberg coherence time on the scale of a few microseconds over several days. 
These fluctuations are attributed to slow drifts in system parameters, which are not considered a fundamental limitation.


\begin{thebibliography}{10}

\bibitem{bakr2009quantum}
W.~S. Bakr, J.~I. Gillen, A. Peng, S. F{\"o}lling, and M. Greiner, A quantum gas microscope for detecting single atoms in a Hubbard-regime optical lattice, {\it Nature} {\bf 462},  74  (2009).

\bibitem{sherson2010single}
J.~F. Sherson, C. Weitenberg, M. Endres, M. Cheneau, I. Bloch, and S. Kuhr, Single-atom-resolved fluorescence imaging of an atomic Mott insulator, {\it Nature} {\bf 467},  68  (2010).

\bibitem{endres2016atom}
M. Endres, H. Bernien, A. Keesling, H. Levine, E.~R. Anschuetz, A. Krajenbrink, C. Senko, V. Vuletic, M. Greiner, and M.~D. Lukin, Atom-by-atom assembly of defect-free one-dimensional cold atom arrays, {\it Science} {\bf 354},  1024  (2016).

\bibitem{barredo2016atom}
D. Barredo, S. De~L{\'e}s{\'e}leuc, V. Lienhard, T. Lahaye, and A. Browaeys, An atom-by-atom assembler of defect-free arbitrary two-dimensional atomic arrays, {\it Science} {\bf 354},  1021  (2016).

\bibitem{norcia2018microscopic}
M. Norcia, A. Young, and A. Kaufman, Microscopic control and detection of ultracold strontium in optical-tweezer arrays, {\it Physical Review X} {\bf 8},  041054  (2018).

\bibitem{cooper2018alkaline}
A. Cooper, J.~P. Covey, I.~S. Madjarov, S.~G. Porsev, M.~S. Safronova, and M. Endres, Alkaline-earth atoms in optical tweezers, {\it Physical Review X} {\bf 8},  041055  (2018).

\bibitem{saskin2019narrow}
S. Saskin, J. Wilson, B. Grinkemeyer, and J. Thompson, Narrow-line cooling and imaging of ytterbium atoms in an optical tweezer array, {\it Physical review letters} {\bf 122},  143002  (2019).

\bibitem{liu2018building}
L. Liu, J. Hood, Y. Yu, J. Zhang, N. Hutzler, T. Rosenband, and K.-K. Ni, Building one molecule from a reservoir of two atoms, {\it Science} {\bf 360},  900  (2018).

\bibitem{anderegg2019optical}
L. Anderegg, L.~W. Cheuk, Y. Bao, S. Burchesky, W. Ketterle, K.-K. Ni, and J.~M. Doyle, An optical tweezer array of ultracold molecules, {\it Science} {\bf 365},  1156  (2019).

\bibitem{kumar2018sorting}
A. Kumar, T.-Y. Wu, F. Giraldo, and D.~S. Weiss, Sorting ultracold atoms in a three-dimensional optical lattice in a realization of Maxwell’s demon, {\it Nature} {\bf 561},  83  (2018).

\bibitem{lee2017defect-free}
W. Lee, H. Kim, and J. Ahn, Defect-free atomic array formation using the Hungarian matching algorithm, {\it Physical Review A} {\bf 95},  053424  (2017).

\bibitem{norcia2019seconds}
M.~A. Norcia, A.~W. Young, W.~J. Eckner, E. Oelker, J. Ye, and A.~M. Kaufman, Seconds-scale coherence on an optical clock transition in a tweezer array, {\it Science} {\bf 366},  93  (2019).

\bibitem{madjarov2019Atomic}
I.~S. Madjarov, A. Cooper, A.~L. Shaw, J.~P. Covey, V. Schkolnik, T.~H. Yoon, J.~R. Williams, and M. Endres, An Atomic-Array Optical Clock with Single-Atom Readout, {\it Phys. Rev. X} {\bf 9},  041052  (2019).

\bibitem{young2020half}
A.~W. Young, W.~J. Eckner, W.~R. Milner, D. Kedar, M.~A. Norcia, E. Oelker, N. Schine, J. Ye, and A.~M. Kaufman, Half-minute-scale atomic coherence and high relative stability in a tweezer clock, {\it Nature} {\bf 588},  408  (2020).

\bibitem{isenhower2010demonstration}
L. Isenhower, E. Urban, X. Zhang, A. Gill, T. Henage, T.~A. Johnson, T. Walker, and M. Saffman, Demonstration of a neutral atom controlled-NOT quantum gate, {\it Physical review letters} {\bf 104},  010503  (2010).

\bibitem{scholl2021quantum}
P. Scholl {\it et~al.}, Quantum simulation of 2D antiferromagnets with hundreds of Rydberg atoms, {\it Nature} {\bf 595},  233  (2021).

\bibitem{ebadi2021quantum}
S. Ebadi {\it et~al.}, Quantum phases of matter on a 256-atom programmable quantum simulator, {\it Nature} {\bf 595},  227  (2021).

\bibitem{levine2019parallel}
H. Levine {\it et~al.}, Parallel implementation of high-fidelity multiqubit gates with neutral atoms, {\it Physical review letters} {\bf 123},  170503  (2019).

\bibitem{madjarov2020high}
I.~S. Madjarov, J.~P. Covey, A.~L. Shaw, J. Choi, A. Kale, A. Cooper, H. Pichler, V. Schkolnik, J.~R. Williams, and M. Endres, High-fidelity entanglement and detection of alkaline-earth Rydberg atoms, {\it Nature Physics} {\bf 16},  857  (2020).

\bibitem{bluvstein2024logical}
D. Bluvstein {\it et~al.}, Logical quantum processor based on reconfigurable atom arrays, {\it Nature} {\bf 626},  58  (2024).

\bibitem{saffman2016quantum}
M. Saffman, Quantum computing with atomic qubits and Rydberg interactions: progress and challenges, {\it Journal of Physics B: Atomic, Molecular and Optical Physics} {\bf 49},  202001  (2016).

\bibitem{de2018analysis}
S. De~L{\'e}s{\'e}leuc, D. Barredo, V. Lienhard, A. Browaeys, and T. Lahaye, Analysis of imperfections in the coherent optical excitation of single atoms to Rydberg states, {\it Physical Review A} {\bf 97},  053803  (2018).

\bibitem{evered2023high}
S.~J. Evered {\it et~al.}, High-fidelity parallel entangling gates on a neutral-atom quantum computer, {\it Nature} {\bf 622},  268  (2023).

\bibitem{manetsch2024tweezer}
H.~J. Manetsch, G. Nomura, E. Bataille, K.~H. Leung, X. Lv, and M. Endres, A tweezer array with 6100 highly coherent atomic qubits, {\it arXiv preprint arXiv:2403.12021}  (2024).

\bibitem{dubielzig2021cryo}
T. Dubielzig, S. Halama, H. Hahn, G. Zarantonello, M. Niemann, A. Bautista-Salvador, and C. Ospelkaus, {Ultra-low-vibration closed-cycle cryogenic surface-electrode ion trap apparatus}, {\it Review of Scientific Instruments} {\bf 92},  043201  (2021).

\bibitem{Labaziewicz2008a}
J. Labaziewicz, Y. Ge, D.~R. Leibrandt, S.~X. Wang, R. Shewmon, and I.~L. Chuang, Temperature {{Dependence}} of {{Electric Field Noise}} above {{Gold Surfaces}}, {\it Physical Review Letters} {\bf 101},  180602  (2008).

\bibitem{Chiaverini2014}
J. Chiaverini and J.~M. Sage, Insensitivity of the Rate of Ion Motional Heating to Trap-Electrode Material over a Large Temperature Range, {\it Physical Review A} {\bf 89},  012318  (2014).

\bibitem{Sedlacek2018a}
J.~A. Sedlacek, J. Stuart, D.~H. Slichter, C.~D. Bruzewicz, R. McConnell, J.~M. Sage, and J. Chiaverini, Evidence for Multiple Mechanisms Underlying Surface Electric-Field Noise in Ion Traps, {\it Physical Review A} {\bf 98},  063430  (2018).

\bibitem{brandl2016cryogenic}
M. Brandl {\it et~al.}, Cryogenic setup for trapped ion quantum computing, {\it Review of Scientific Instruments} {\bf 87},  113103  (2016).

\bibitem{pagano2018cryogenic}
G. Pagano {\it et~al.}, Cryogenic trapped-ion system for large scale quantum simulation, {\it Quantum Science and Technology} {\bf 4},  014004  (2018).

\bibitem{Todaro2020}
S.~L. Todaro, Improved {{State Detection}} and {{Transport}} of {{Trapped Ion Qubits}} for {{Scalable Quantum Computing}}, Ph.D. thesis, University of Colorado at Boulder, 2020.

\bibitem{raimond2001manipulating}
J.-M. Raimond, M. Brune, and S. Haroche, Manipulating quantum entanglement with atoms and photons in a cavity, {\it Reviews of Modern Physics} {\bf 73},  565  (2001).

\bibitem{nguyen2018towards}
T.~L. Nguyen {\it et~al.}, Towards quantum simulation with circular Rydberg atoms, {\it Physical Review X} {\bf 8},  011032  (2018).

\bibitem{cantat2020long}
T. Cantat-Moltrecht, R. Corti{\~n}as, B. Ravon, P. M{\'e}haignerie, S. Haroche, J.-M. Raimond, M. Favier, M. Brune, and C. Sayrin, Long-lived circular Rydberg states of laser-cooled rubidium atoms in a cryostat, {\it Physical Review Research} {\bf 2},  022032  (2020).

\bibitem{cohen2021quantum}
S.~R. Cohen and J.~D. Thompson, Quantum computing with circular Rydberg atoms, {\it PRX Quantum} {\bf 2},  030322  (2021).

\bibitem{semeghini2021probing}
G. Semeghini {\it et~al.}, Probing topological spin liquids on a programmable quantum simulator, {\it Science} {\bf 374},  1242  (2021).

\bibitem{zeiher2017realization}
J. Zeiher, Realization of Rydberg-dressed quantum magnets, Ph.D. thesis, Ludwig Maximilians Universit{\"a}t M{\"u}nchen, 2017.

\bibitem{beterov2009quasiclassical}
I. Beterov, I. Ryabtsev, D. Tretyakov, and V. Entin, Quasiclassical calculations of blackbody-radiation-induced depopulation rates and effective lifetimes of Rydberg n S, n P, and n D alkali-metal atoms with $n \leq 80$, {\it Physical review A} {\bf 79},  052504  (2009).

\bibitem{wu2022erasure}
Y. Wu, S. Kolkowitz, S. Puri, and J.~D. Thompson, Erasure conversion for fault-tolerant quantum computing in alkaline earth Rydberg atom arrays, {\it Nature communications} {\bf 13},  4657  (2022).

\bibitem{ma2023high}
S. Ma, G. Liu, P. Peng, B. Zhang, S. Jandura, J. Claes, A.~P. Burgers, G. Pupillo, S. Puri, and J.~D. Thompson, High-fidelity gates and mid-circuit erasure conversion in an atomic qubit, {\it Nature} {\bf 622},  279  (2023).

\bibitem{scholl2023erasure}
P. Scholl, A.~L. Shaw, R.~B.-S. Tsai, R. Finkelstein, J. Choi, and M. Endres, Erasure conversion in a high-fidelity Rydberg quantum simulator, {\it Nature} {\bf 622},  273  (2023).

\bibitem{chow2024circuit}
M.~N. Chow, V. Buchemmavari, S. Omanakuttan, B.~J. Little, S. Pandey, I.~H. Deutsch, and Y.-Y. Jau, Circuit-Based Leakage-to-Erasure Conversion in a Neutral-Atom Quantum Processor, {\it PRX Quantum} {\bf 5},  040343  (2024).

\bibitem{boulier2017spontaneous}
T. Boulier, E. Magnan, C. Bracamontes, J. Maslek, E. Goldschmidt, J. Young, A.~V. Gorshkov, S. Rolston, and J.~V. Porto, Spontaneous avalanche dephasing in large Rydberg ensembles, {\it Physical Review A} {\bf 96},  053409  (2017).

\bibitem{festa2022blackbody}
L. Festa, N. Lorenz, L.-M. Steinert, Z. Chen, P. Osterholz, R. Eberhard, and C. Gross, Blackbody-radiation-induced facilitated excitation of Rydberg atoms in optical tweezers, {\it Physical Review A} {\bf 105},  013109  (2022).

\bibitem{zeiher2016many}
J. Zeiher, R. Van~Bijnen, P. Schau{\ss}, S. Hild, J.-y. Choi, T. Pohl, I. Bloch, and C. Gross, Many-body interferometry of a Rydberg-dressed spin lattice, {\it Nature Physics} {\bf 12},  1095  (2016).

\bibitem{manovitz2025quantum}
T. Manovitz {\it et~al.}, Quantum coarsening and collective dynamics on a programmable simulator, {\it Nature} {\bf 638},  86  (2025).

\bibitem{safronova2012blackbody}
M.~S. Safronova, M.~G. Kozlov, and C.~W. Clark, Blackbody radiation shifts in optical atomic clocks, {\it IEEE transactions on ultrasonics, ferroelectrics, and frequency control} {\bf 59},  439  (2012).

\bibitem{ushijima2015cryogenic}
I. Ushijima, M. Takamoto, M. Das, T. Ohkubo, and H. Katori, Cryogenic optical lattice clocks, {\it Nature Photonics} {\bf 9},  185  (2015).

\bibitem{beloy2014atomic}
K. Beloy, N. Hinkley, N.~B. Phillips, J.~A. Sherman, M. Schioppo, J. Lehman, A. Feldman, L.~M. Hanssen, C.~W. Oates, and A.~D. Ludlow, Atomic clock with 1$\times$ 10-18 room-temperature blackbody Stark uncertainty, {\it Physical review letters} {\bf 113},  260801  (2014).

\bibitem{ni2018dipolar}
K.-K. Ni, T. Rosenband, and D.~D. Grimes, Dipolar exchange quantum logic gate with polar molecules, {\it Chemical science} {\bf 9},  6830  (2018).

\bibitem{holland2024demonstration}
C.~M. Holland, Y. Lu, S.~J. Li, C.~L. Welsh, and L.~W. Cheuk, Demonstration of Erasure Conversion in a Molecular Tweezer Array, {\it arXiv preprint arXiv:2406.02391}  (2024).

\bibitem{Liu2024}
Y. Liu, J. Schmidt, Z. Liu, D.~R. Leibrandt, D. Leibfried, and C. wen Chou, Quantum state tracking and control of a single molecular ion in a thermal environment, {\it Science} {\bf 385},  790  (2024).

\bibitem{takamoto2011frequency}
M. Takamoto, T. Takano, and H. Katori, Frequency comparison of optical lattice clocks beyond the Dick limit, {\it Nature Photonics} {\bf 5},  288  (2011).

\bibitem{norcia2024iterative}
M. Norcia {\it et~al.}, Iterative Assembly of 171 Yb Atom Arrays with Cavity-Enhanced Optical Lattices, {\it PRX Quantum} {\bf 5},  030316  (2024).

\bibitem{gyger2024continuous}
F. Gyger, M. Ammenwerth, R. Tao, H. Timme, S. Snigirev, I. Bloch, and J. Zeiher, Continuous operation of large-scale atom arrays in optical lattices, {\it Physical Review Research} {\bf 6},  033104  (2024).

\bibitem{tian2023parallel}
W. Tian, W.~J. Wee, A. Qu, B.~J.~M. Lim, P.~R. Datla, V.~P.~W. Koh, and H. Loh, Parallel assembly of arbitrary defect-free atom arrays with a multitweezer algorithm, {\it Physical Review Applied} {\bf 19},  034048  (2023).

\bibitem{bothwell2022resolving}
T. Bothwell, C.~J. Kennedy, A. Aeppli, D. Kedar, J.~M. Robinson, E. Oelker, A. Staron, and J. Ye, Resolving the gravitational redshift across a millimetre-scale atomic sample, {\it Nature} {\bf 602},  420  (2022).

\bibitem{zheng2022differential}
X. Zheng, J. Dolde, V. Lochab, B.~N. Merriman, H. Li, and S. Kolkowitz, Differential clock comparisons with a multiplexed optical lattice clock, {\it Nature} {\bf 602},  425  (2022).

\bibitem{clements2020lifetime}
E.~R. Clements, M.~E. Kim, K. Cui, A.~M. Hankin, S.~M. Brewer, J. Valencia, J.-S. Chen, C.-W. Chou, D.~R. Leibrandt, and D.~B. Hume, Lifetime-limited interrogation of two independent al+ 27 clocks using correlation spectroscopy, {\it Physical review letters} {\bf 125},  243602  (2020).

\bibitem{timmermans2001degenerate}
E. Timmermans, Degenerate fermion gas heating by hole creation, {\it Physical Review Letters} {\bf 87},  240403  (2001).

\bibitem{ji2024observation}
Y. Ji, J. Chen, G.~L. Schumacher, G.~G. Assump{\c{c}}{\~a}o, S. Huang, F.~J. Vivanco, and N. Navon, Observation of the Fermionic Joule-Thomson Effect, {\it Physical Review Letters} {\bf 132},  153402  (2024).

\bibitem{Schymik2021single}
K.-N. Schymik, S. Pancaldi, F. Nogrette, D. Barredo, J. Paris, A. Browaeys, and T. Lahaye, Single Atoms with 6000-Second Trapping Lifetimes in Optical-Tweezer Arrays at Cryogenic Temperatures, {\it Phys. Rev. Applied} {\bf 16},  034013  (2021).

\bibitem{pichard2024rearrangement}
G. Pichard {\it et~al.}, Rearrangement of individual atoms in a 2000-site optical-tweezer array at cryogenic temperatures, {\it Phys. Rev. Appl.} {\bf 22},  024073  (2024).

\bibitem{blodgett2023imaging}
K.~N. Blodgett, D. Peana, S.~S. Phatak, L.~M. Terry, M.~P. Montes, and J.~D. Hood, Imaging a Li 6 Atom in an Optical Tweezer 2000 Times with $\Lambda$-Enhanced Gray Molasses, {\it Physical Review Letters} {\bf 131},  083001  (2023).

\bibitem{norrgard2021quantum}
E.~B. Norrgard, S.~P. Eckel, C.~L. Holloway, and E.~L. Shirley, Quantum blackbody thermometry, {\it New Journal of Physics} {\bf 23},  033037  (2021).

\bibitem{mamat2024mitigating}
B. Mamat {\it et~al.}, Mitigating the noise of residual electric fields for single Rydberg atoms with electron photodesorption, {\it Physical Review Applied} {\bf 22},  064021  (2024).

\bibitem{thiele2014manipulating}
T. Thiele, S. Filipp, J.~A. Agner, H. Schmutz, J. Deiglmayr, M. Stammeier, P. Allmendinger, F. Merkt, and A. Wallraff, Manipulating Rydberg atoms close to surfaces at cryogenic temperatures, {\it Physical Review A} {\bf 90},  013414  (2014).

\bibitem{vittorini2013modular}
G. Vittorini, K. Wright, K.~R. Brown, A.~W. Harter, and S.~C. Doret, Modular cryostat for ion trapping with surface-electrode ion traps, {\it Review of Scientific Instruments} {\bf 84},  043112  (2013).

\bibitem{hsu2024high}
T.-W. Hsu, A High Optical Access Cryogenic Optical Tweezer Array, Ph.D. thesis, University of Colorado at Boulder, 2024.

\bibitem{ram2011note}
S. Ram, S. Mishra, S. Tiwari, and S. Mehendale, Note: Investigation of atom transfer using a red-detuned push beam in a double magneto-optical trap setup, {\it Review of Scientific Instruments} {\bf 82},  126108  (2011).

\bibitem{bruneau2014guided}
Y. Bruneau, G. Khalili, P. Pillet, and D. Comparat, Guided and focused slow atomic beam from a 2 dimensional magneto optical trap, {\it The European Physical Journal D} {\bf 68},  1  (2014).

\bibitem{willems1995creating}
P. Willems and K. Libbrecht, Creating long-lived neutral-atom traps in a cryogenic environment, {\it Physical Review A} {\bf 51},  1403  (1995).

\bibitem{weiner1999experiments}
J. Weiner, V.~S. Bagnato, S. Zilio, and P.~S. Julienne, Experiments and theory in cold and ultracold collisions, {\it Reviews of Modern Physics} {\bf 71},  1  (1999).

\bibitem{dalibard1989laser}
J. Dalibard and C. Cohen-Tannoudji, Laser cooling below the Doppler limit by polarization gradients: simple theoretical models, {\it JOSA B} {\bf 6},  2023  (1989).

\bibitem{brown2019gray}
M. Brown, T. Thiele, C. Kiehl, T.-W. Hsu, and C. Regal, Gray-molasses optical-tweezer loading: controlling collisions for scaling atom-array assembly, {\it Physical Review X} {\bf 9},  011057  (2019).

\bibitem{jenkins2022ytterbium}
A. Jenkins, J.~W. Lis, A. Senoo, W.~F. McGrew, and A.~M. Kaufman, Ytterbium nuclear-spin qubits in an optical tweezer array, {\it Physical Review X} {\bf 12},  021027  (2022).

\bibitem{brown2001interval}
L.~D. Brown, T.~T. Cai, and A. DasGupta, Interval estimation for a binomial proportion, {\it Statistical science} {\bf 16},  101  (2001).

\bibitem{tuchendler2008energy}
C. Tuchendler, A.~M. Lance, A. Browaeys, Y.~R. Sortais, and P. Grangier, Energy distribution and cooling of a single atom in an optical tweezer, {\it Physical Review A} {\bf 78},  033425  (2008).

\bibitem{ang2022gray}
J. Ang'Ong'A, C. Huang, J.~P. Covey, and B. Gadway, Gray molasses cooling of K 39 atoms in optical tweezers, {\it Physical Review Research} {\bf 4},  013240  (2022).

\bibitem{hutzler2017eliminating}
N.~R. Hutzler, L.~R. Liu, Y. Yu, and K.-K. Ni, Eliminating light shifts for single atom trapping, {\it New Journal of Physics} {\bf 19},  023007  (2017).

\bibitem{covey20192000}
J.~P. Covey, I.~S. Madjarov, A. Cooper, and M. Endres, 2000-times repeated imaging of strontium atoms in clock-magic tweezer arrays, {\it Physical review letters} {\bf 122},  173201  (2019).

\bibitem{schymik2022scaling}
K.-N. Schymik, Scaling-up the Tweezer Platform-Trapping Arrays of Single Atoms in a Cryogenic Environment, Ph.D. thesis, Universit{\'e} Paris-Saclay, 2022.

\bibitem{efron1994introduction}
B. Efron and R.~J. Tibshirani, {\it An introduction to the bootstrap} (Chapman and Hall/CRC, London, 1994).

\bibitem{klein2003survival}
J.~P. Klein {\it et~al.}, {\it Survival analysis: techniques for censored and truncated data} (Springer, New York, 2003), Vol.~1230.

\bibitem{arpornthip2012vacuum}
T. Arpornthip, C. Sackett, and K. Hughes, Vacuum-pressure measurement using a magneto-optical trap, {\it Physical Review A} {\bf 85},  033420  (2012).

\bibitem{barker2022precise}
D.~S. Barker, B.~P. Acharya, J.~A. Fedchak, N.~N. Klimov, E.~B. Norrgard, J. Scherschligt, E. Tiesinga, and S.~P. Eckel, Precise quantum measurement of vacuum with cold atoms, {\it Review of Scientific Instruments} {\bf 93},    (2022).

\bibitem{wang2023accelerating}
S. Wang, W. Zhang, T. Zhang, S. Mei, Y. Wang, J. Hu, and W. Chen, Accelerating the assembly of defect-free atomic arrays with maximum parallelisms, {\it Physical Review Applied} {\bf 19},  054032  (2023).

\bibitem{shadmany2025cavity}
D. Shadmany {\it et~al.}, Cavity QED in a high NA resonator, {\it Science Advances} {\bf 11},  eads8171  (2025).

\bibitem{hu2025site}
B. Hu, J. Sinclair, E. Bytyqi, M. Chong, A. Rudelis, J. Ramette, Z. Vendeiro, and V. Vuleti{\'c}, Site-selective cavity readout and classical error correction of a 5-bit atomic register, {\it Physical Review Letters} {\bf 134},  120801  (2025).

\bibitem{bernien2017probing}
H. Bernien {\it et~al.}, Probing many-body dynamics on a 51-atom quantum simulator, {\it Nature} {\bf 551},  579  (2017).

\bibitem{levine2018high}
H. Levine, A. Keesling, A. Omran, H. Bernien, S. Schwartz, A.~S. Zibrov, M. Endres, M. Greiner, V. Vuleti{\'c}, and M.~D. Lukin, High-fidelity control and entanglement of Rydberg-atom qubits, {\it Physical review letters} {\bf 121},  123603  (2018).

\bibitem{graham2019rydberg}
T. Graham, M. Kwon, B. Grinkemeyer, Z. Marra, X. Jiang, M. Lichtman, Y. Sun, M. Ebert, and M. Saffman, Rydberg-mediated entanglement in a two-dimensional neutral atom qubit array, {\it Phys. Rev. Lett.} {\bf 123},  230501  (2019).

\bibitem{martinez2018state}
M. Martinez-Dorantes, W. Alt, J. Gallego, S. Ghosh, L. Ratschbacher, and D. Meschede, State-dependent fluorescence of neutral atoms in optical potentials, {\it Physical Review A} {\bf 97},  023410  (2018).

\bibitem{stricker2020experimental}
R. Stricker, D. Vodola, A. Erhard, L. Postler, M. Meth, M. Ringbauer, P. Schindler, T. Monz, M. M{\"u}ller, and R. Blatt, Experimental deterministic correction of qubit loss, {\it Nature} {\bf 585},  207  (2020).

\bibitem{eckner2023realizing}
W.~J. Eckner, N. Darkwah~Oppong, A. Cao, A.~W. Young, W.~R. Milner, J.~M. Robinson, J. Ye, and A.~M. Kaufman, Realizing spin squeezing with Rydberg interactions in an optical clock, {\it Nature} {\bf 621},  734  (2023).

\end{thebibliography}
\end{document}